\documentclass{aa}

\usepackage[dvips]{graphicx}
\usepackage{txfonts}
\usepackage{longtable,lscape}
\usepackage{natbib}
\usepackage{url}
\usepackage{subfig}
\bibpunct{(}{)}{;}{a}{}{,}
\usepackage{color}
\usepackage{url}

\begin{document}
   \title{The Besan\c{c}on Galaxy model renewed} 
 \subtitle{I. Constraints on the local star formation history from Tycho data }
 \author{ M. A. Czekaj \inst{\ref{inst1}}
      \and A. C. Robin \inst{\ref{inst2}}
      \and F. Figueras \inst{\ref{inst1}}
      \and X. Luri \inst{\ref{inst1}}
      \and M. Haywood \inst{\ref{inst3}}
          }
   \offprints{M. A. Czekaj,
   \email{mczekaj@am.ub.es}}
   \institute{Departament d'Astronomia i Meteorologia and IEEC-ICC-UB,
     Universitat de Barcelona,
     Mart\'i i Franqu\`es, 1, E-08028 Barcelona, Spain \label{inst1} \and
     Institute Utinam, CNRS UMR6213, Universit\'e de Franche-Comt\'e, OSU THETA de Franche-Comt\'e-Bourgogne, Besan\c{c}on, France \label{inst2}\and
     GEPI, Observatoire de Paris, CNRS, Universit\'e Paris Diderot, 92190, Meudon, France \label{inst3}}
   \date{Received  / Accepted }
 
\abstract
{The understanding of Galaxy evolution can be facilitated by the use of population synthesis models, which allows us
to test hypotheses on the star formation history, star evolution, and chemical and dynamical evolution of the 
Galaxy. } 
{The new version of the Besan\c{c}on Galaxy model (hereafter BGM) aims to provide a more flexible and powerful tool 
to investigate the initial mass function (IMF) and star formation rate (SFR) of the Galactic disc.}
{We present a new strategy for the generation of thin disc stars, which assumes the IMF, SFR and evolutionary
tracks as free parameters. We have updated most of the ingredients for the star count production and, for the first
time, binary stars are generated in a consistent way. The local dynamical self-consistency is maintained in this 
new scheme.
We then compare simulations from the new model with Tycho-2 data and the local luminosity
function, as a first test to verify and constrain the new ingredients. The effects of changing thirteen different
 ingredients of the model are systematically studied.}
{For the first time, a full sky comparison is performed between BGM and data. This strategy allows us to constrain the IMF 
slope at high masses, which is found to be close to 3.0 and excludes a shallower slope such as Salpeter's one. 
The SFR is found decreasing whatever IMF is assumed. 
The model is compatible with a local dark matter density of 0.011 $M_\odot$pc$^{-3}$ implying that there is no compelling
 evidence for the significant amount of dark matter 
in the disc. While the model is fitted to Tycho-2 data, which is a magnitude limited sample with V$<$11, we check that it is still 
consistent with fainter stars.}
{The new model constitutes a new basis for further comparisons with large scale surveys and is being prepared
to become a powerful tool for the analysis of the Gaia mission data.}


\maketitle

\section{Introduction}\label{intro}

The understanding of the origin and evolution of the Milky Way is one of the primary goals of 
the Gaia mission (ESA, launched December 2013). The use of its data 
to test different hypotheses and scenarios of galaxy formation and evolution requires the 
availability of an adaptable Galaxy model to provide simulated data for comparison. Kinematic 
and photometric data with the ages and metallicities of the stars and their statistics allows us
 to characterise galaxy's populations and, from that, the 
overall Galactic gravitational potential.  One of the most promising procedures to reach such a
goal is to optimize the present population synthesis models by fitting robust statistical 
techniques, the large and small scale structure, and kinematics parameters that best reproduce 
Gaia data. The work presented in this paper is focused on the optimization of the structure parameters of 
the Milky Way Galactic disc in the BGM by comparing the simulations to real data to study the process of Galaxy evolution.

The development of Galaxy models started in the early 80's when several needs for star count predictions for the sake
of preparing observations, in particular HST mission. For example, \cite{Bahcall} built a model based on simple 
assumptions on density laws and luminosity functions, while \cite{Cohen} developed a model in the infrared using a more
complex set of luminosity functions but leading to a very large number of free parameters. From the beginning, the
aim of the BGM was not only to be able to simulate reasonable star counts but 
further to test scenarios of Galactic evolution from assumptions on the rate of star formation, initial mass 
function, and stellar evolution. This was the spirit of the \cite{Tinsley1972A&A....20..383T} formalism of decomposition
of the star-creation process in these two functions of mass and time, which are assumed independent. 
\cite{Creze1979} and \cite{CrezeRobin1983} conceived the starting point of a model to
be able to generate 
samples of stars directly comparable with observations. The idea was that simulations can be efficiently compared with
data because all observational bias can easily be taken into account. Then statistical tests and fitting processes can
be applied to constrain model parameters and hypothesis, such as the IMF, the density laws, etc. The initial BGM 
from \cite{ar1986} was improved for taking into account dynamical constraints (see \cite{bienayme}).
Further developments 
concerned the stellar and Galactic evolution constraints from the Hipparcos input catalogue, the high latitude star counts
(\cite{misha1997-2}), constraints on the thick disc population \citep{Robin1996,Reyle2001},
 the halo (\cite{Robin2000}), and the bulge \citep{Picaud2004}. 
This leads to a version, which was released to the community on the web \footnote{http://model.obs-besancon.fr} and 
described in \cite{ar2003}. 
Since this work, we continued to develop it, partly for giving improved fit to available data and partly to constrain 
Galactic structure and scenarios of galaxy formation, especially in the Galactic plane, thanks to the 3D extinction model 
developed \citep{marshall3D2006}, the study of the Galactic warp and flare \citep{Derriere,Reyle2009}, and, lastly, the bar 
and bulge regions \citep{ar2012}.

In recent years, other population synthesis models have been developed, such as those by \cite{NG}, \cite{Vallenari},
 \cite{Girardi2005}, and that known as tri-dimensional model of the galaxy (TRILEGAL). These models are also
based on stellar evolutionary tracks and assumptions on SFH and IMF. However, none of them have 
yet used dynamical constraints, as BGM has.
 The model TRILEGAL benefits for a wide coverage of all the photometric systems used in observations. It uses slightly 
 different model parameters; the atmosphere model used are the one from Padova, while we use semi-empirical
 grids from BaSeL2.2 \citep{Lejeune1997,Lejeune1998} or 3.1 \citep{Westera2002} complemented at low temperatures by NextGen
 models in BGM. The colour
 predictions are then significantly different in the low mass region. 
 The TRILEGAL model uses essentially the same kind of parametrization for the thin disc with a scale height changing with
 age but with an exponential or secant squared while BGM uses Einasto ellipsoids. A simple diffuse extinction model from
 an exponential or sech$^{2}$ function is used for modelling the interstellar absorption, and the warp and flare in the outer
 disc are not yet taken into account in TRILEGAL. Bulge parameters, as proposed by \cite{vanhollobeke} are sensitively
 different than in BGM \citep{ar2012}.
 \cite{NG} and \cite{Vallenari} models have only been compared to data in a limited number of 
 directions of observations, and their parameters are less well established and constrained. 
 
\cite{JJ2010} proposed a new scheme, using local dynamical constraints to create a local model of the Galaxy.
This model only accounts for main sequence stars and have not been compared with large scale surveys, but it produced 
interesting constraints on the local star formation history and on the IMF.
 \cite{Gao} attempted to combine the \cite{JJ2010} approach with TRILEGAL code to fit some parameters and 
 improved the fit to SDSS data towards high latitudes. However, they conclude that TRILEGAL has not enough flexibility
 to obtain a satisfactory fit to these data and show that BGM (2003 version) does not compare well with SDSS data. 
 However, this was a version of the BGM that was not calibrated for the ugriz photometric system yet.

Galaxia (\cite{Sharma}) is another type of approach, which makes use of the BGM (the 2003 version) in 
order to create large simulations to be compared with large scale surveys. It can take any Galaxy model 
as an input but is not a new Galaxy model by itself. It is also a very useful tool to translate N-body 
simulations into an observable catalogue.
 
For a long time, we have identified a systematic disagreement between BGM and real data concerning bright stars. A 
comparison to the Hipparcos catalogue showed that the A-F dwarfs are over-represented by 20-30\% while the quality of
star count predictions are at the level of a few percent at fainter magnitudes (\cite{ar2003,IbataI}). It appears that 
faint star counts (typically  V$\ge$ 15) are dominated by thick disc and halo populations at or slightly below the turnoff, 
while the contribution of the thin disc is dominated by K and M dwarfs, a region of the HR diagram where the
history of star formation is not important. With these stars being unevolved, this occurs because their density depends on 
the integral of the SFR over
the age of the disc and not on the detailed history of the SFR.  Since the study of \cite{misha1997-1} on the Hipparcos 
input catalogue, which showed that the SFR should be nearly constant or not varying by more than a factor of 3 over 
the age of the thin disc, we assumed a constant star formation for the thin disc in BGM. The A-F dwarfs, 
which constitute a large contribution to bright data sets, as Hipparcos and Tycho-2, are indeed very sensitive to the 
history of the SFR, and we concluded that it was necessary to specifically study this aspect using 
these dwarfs in the solar neighbourhood to get rid of the discrepancy of the BGM. In doing so, the Tycho-2 catalogue is 
the best homogeneous data set available for the whole sky and complete to about magnitude $V \sim 11$. This is the specific
subject of the present study.

In the era of Gaia, a model able to analyse stellar population distributions to constrain Galaxy evolution 
would be very welcome. The BGM is in this respect a very useful tool and should be able to ``measure'' the star 
formation history in different regions of the Milky Way. However, the old scheme of BGM, based on the evolution 
scheme from \cite{misha1997-1} does not have enough flexibility to change the IMF and SFR in the thin disc. Moreover,
it does not allow us to generate binary stars easily. This is why we have undertaken the present work to modify the BGM 
code to be able to change these parameters easily to use different ones in different regions of the Milky Way if 
necessary and to account for stellar binarity at the same time. To test this scheme and to constrain the 
 star formation history, the new code presented here has been compared and tested using the Tycho-2 catalogue and the
local luminosity function (LF). We consider this as a first 
step for furnishing a reliable model for interpretation of future large scale surveys, such as Gaia, but also for RAVE, 
APOGEE, Gaia-ESO, and other future data.

 In Section \ref{bgmCode}, the new 
approach and the overall structure of the code are presented 
. In Section 
\ref{tools}, we present the observables and tools used 
in our analysis. In Section \ref{sec:ingred}, we detail the model ingredients, which have been 
investigated. In Section \ref{res}, we show the results obtained from the
multiple comparisons of our model with the Tycho-2 data and the LF; additionally it is our first check of how the new model 
reproduces deep star counts.

\section{The new version of Besan\c{c}on Galaxy model}\label{bgmCode}
In this work, we have focused on the Galactic thin disc; thus, the developments presented here concern
the treatment of that population only. The thin disc population represents $\sim$ 92-98 \% of the sample used in the present 
analysis.  Work is in progress to apply the same scheme to other populations. For the sake of comparison to Tycho
data, the other components, the thick disc, and the halo are simulated by single bursts of star formation, as in 
\cite{ar2003},
and the corresponding Hess diagrams are used, where the bulge is negligible locally.

\subsection{A new approach to the BGM}\label{sec:newModel}

\subsubsection{The overall structure}

The version of the BGM presented here is based on a new approach to the star simulation procedure.
Rather than using a fixed Hess diagram, we have turned the IMF, SFR, and evolutionary 
tracks into free parameters. The model performs the star-creation process 
assuming an IMF, a SFR along the age of the Galactic disc, and an age-metallicity 
relation. Created stars follow a set of evolutionary tracks and may finish as remnants. To 
generate a simulated Galaxy from these fundamental buildingblocks, we have implemented important changes 
in the BGM code.

\begin{figure}
\begin{center}
\includegraphics[width=9cm]{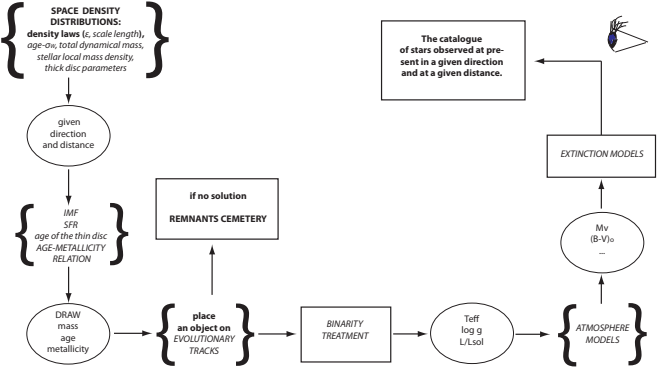}\\
\end{center}
\caption{General scheme describing the ingredients needed to build the evolutionary model and the information
 they provide. In italics we have marked the ingredients, which are under study, in this paper.}
\label{ingredients} 
\end{figure}

Figure \ref{ingredients} presents the scheme of the new model's ingredients and the information they 
provide. First, we draw the mass, the age, and the metallicity of an object, according to the chosen 
functions. With the obtained values we search the location of the star in the HR diagram by interpolating the evolutionary tracks. 
If such a solution is found, then a star with the given $L/L_{\odot}$, log g, and Teff is created. Using the 
atmosphere models, we then compute its observed parameters and include the extinction effects and optionally 
the estimations of the observational errors.
From the {\it a priori} knowledge on the space density distributions, we know how much mass is to
 be converted 
into stars at various distances within the Galactic disc. Combining all this information, we get 
the picture of the thin disc at the present time. The stars for which a solution on the tracks was not found, 
meaning that 
the combination of their age and mass does not correspond to an alive star, are moved to the ``remnants 
cemetery''. 

A still missing feature in this new thin disc treatment is a correct treatment of white dwarfs, which we 
are not yet producing in a consistent way using their evolutionary tracks. This is left for future work. At the 
moment, they are produced following the scheme of the old model.

\subsubsection{Determination of the mass model}\label{sec:massModel}

As explained in \citet{ar1986}, the definition of the mass density $\rho(r,l,b,i)$ at heliocentric distance $r$ in the direction
of the galactocentric coordinates $(l,b)$ for each age subcomponent $i$ requires the following:

\begin{enumerate}
 \item an estimation of the mass density of each subcomponent at the Sun position,
 \item a mathematical law capable of reproducing the trend of the density from the solar neighborhood 
       to remote distances, and
 \item evaluations of both the scale heights and the scale lengths of each subcomponent to constrain free 
       parameters of the mathematical expression.
\end{enumerate}

As explained in \cite{ar2003}, the BGM thin disc is divided into seven age subcomponents. 
This division is maintained in the new model and could be easily changed to a larger number of subcomponents if 
required. Each supcomponent is modelled by the Einasto density law, 
as seen in Table 3 of \cite{ar2003}. It was chosen to represent the thin disc because it assures the continuity and 
derivability in the plane.  The Einasto law ellipsoids are 
defined by a scale length and an eccentricity $\epsilon$. The eccentricities for all disc subcomponents are 
obtained from dynamic considerations, as explained later in this section. Applying the Einasto law, we calculate the
relative density
at the $(x, y, z)$ position. Depending on the age subcomponent and the user specifications, it may include the warp, 
flare, and spiral arms.

\begin{figure*}[ht!]
\small
\setlength{\unitlength}{1cm}
\begin{picture}(18,13)(-2,10)

%
%
{\thicklines
\put(-2,20.5){\framebox(18,2.5)}
}

\put(-1.9,21.7){Model ingredients:}

\put(1.3,21){\framebox(3,1.5){$\begin{array}{c}
                           \mbox{SFR(i), thin disc age, } \\
                           \mbox{$\rho_{\odot}^{obs}$ from LF} \\

                           \end{array}$}}

\put(5.2,21){\framebox(5,1.5){$\begin{array}{c}
                           \mbox{IMF, evolutionary tracks,} \\
                           \mbox{age-metallicity relation, binarity} \\
                           \end{array}$}}

\put(11,21){\framebox(3,1.5){$\begin{array}{c}
                           \mbox {age-$\sigma_W$ relation, } \\
                           \mbox {dynamical mass}
                           \end{array}$}}

%
%

\put(-1.9,11){\framebox(5,8)}

\begin{Large}\put(-1.8,19.3){A}\end{Large}
\put(-1.6,19.5){\circle{0.8}}

\put(-1.65,18.4){LOCAL NORMALIZATION}
\put(-1.8,17.4){Inputs: disc eccentricities $\epsilon(i)$, $\Sigma_{\odot}^{all}$}
\put(-1.5,16.8){~~~~~~~~~$\rho_{\odot}^{all} = \rho_{\odot}^{obs} + \rho_{\odot}^{remnants}$, SFR(i)}
\put(-1.8,15.5){Conditions:}
 \put(-1.2,14.7){1. $\Sigma_{\odot}^{all}(i) \sim \Sigma_{\odot}^{all} \times SFR(i)$}
\put(-0.6,13.3){2. $\rho_{\odot}^{all} = \sum\limits_{1}^{7}  \rho_{\odot}^{all}(i)$}
\put(-1.8,11.7){Output:  $~~~~~\rho_{\odot}^{all}(i)$}
\put(3.1,11.7){\line(1,0){0.5}}
\put(3.6,11.7){\line(0,1){5.8}}
\put(3.6,17.53){\vector(1,0){0.5}}

\put(4.1,11){\framebox(5,8)}

\begin{Large}\put(4.45,19.3){B}\end{Large}
\put(4.6,19.5){\circle{0.8}}

\put(4.15,18.4){SPHERE AROUND THE SUN}
\put(4.15,17.4){Inputs: $~~~~~~\rho_{\odot}^{all}(i)$}
\put(4.5,16){ Using the new model we}
\put(4.5,15.5){ simulate a sphere around}
\put(4.3,15){ the Sun. Stars are produced }
\put(4.5,14.5){according to given input }
\put(4.15,14){ ingredients. Some of them are }
\put(4.3,13.5){ found still alive and others}
\put(4.3,13){ went already to remnants. }
\put(4.15,11.7){Output:  $~~~~~\rho_{\odot}^{obs}(i), ~~\rho_{\odot}^{remnants}(i) $ }
\put(9.1,11.7){\line(1,0){0.5}}
\put(9.6,11.7){\line(0,1){5.8}}
\put(9.6,17.53){\vector(1,0){0.5}}
\put(10.1,11){\framebox(6,8)}

\begin{Large}\put(11.35,19.3){C}\end{Large}
\put(11.5,19.5){\circle{0.8}}
\put(10.2,18.4){DYNAMICAL SELF-CONSISTENCY}
\put(10.4,17.4){Inputs: $~~~\rho_{\odot}^{obs}(i)$ + $\rho_{\odot}^{WD}(i)$ + $\rho_{thick}$}
\put(11.5,16.8){ + $\rho_{halo}$ + $\rho_{bulge}$ + $\rho_{ISM}$}
\put(11.5,16.2){  age-$\sigma_W$ relation}
\put(10.4,15.4){Iteratively solve:}
\put(11.5,14.7){$\nabla^{2}  \Phi (r,z) = 4 \pi  \rho(r,z) G$}
\put(10.6,13.7){$ \sigma_W^2 \ln (\frac{\rho(r,z)}{\rho(r,0)}) = -\Phi(r,z) + \Phi(r,0)$}
\put(10.6,12.7){fitting $\rho_{DH}$ and bulge parameters}
\put(10.4,11.7){Output: $~~~~~\epsilon(i)$, $\Phi$ }
\put(13,11){\line(0,-1){0.9}}
\put(13,10.1){\line(-1,0){15.4}}
\put(-2.4,10.1){\line(0,1){7.5}}
\put(-2.4,17.6){\vector(1,0){0.5}}

{\thicklines
\put(2,21){\vector(0,-1){2}}
}

{\thicklines
\put(6.6,21){\vector(0,-1){2}}
}

{\thicklines
\put(12.6,21){\vector(0,-1){2}}
}
\end{picture}
\caption{Normalization iterative process to derive the mass model.
  The local volume density $\rho_{\odot}^{all}$ is the sum of the stellar volume mass density of 
  alive stars $\rho_{\odot}^{obs}$ (derived from the luminosity function and the mass-luminosity relation) and a 
  fraction representing the volume density of remnants $\rho_{\odot}^{remnants}$. 
  The stellar mass density of each age bin $\rho_{\odot}^{all}(i)$ is derived in block A and then applied in Eq. \ref{reservmass1} to calculate the \textit{mass reservoir}. SFR(i) 
  is a vector that stores seven dimensionless factors normalized to unity, reflecting the intensity of the 
  SFR during each age interval.The  $\Sigma_{\odot}^{all}$ is the total (all ages) surface density 
  in the solar neighborhood (SN).}
\label{massModel}
\end{figure*}
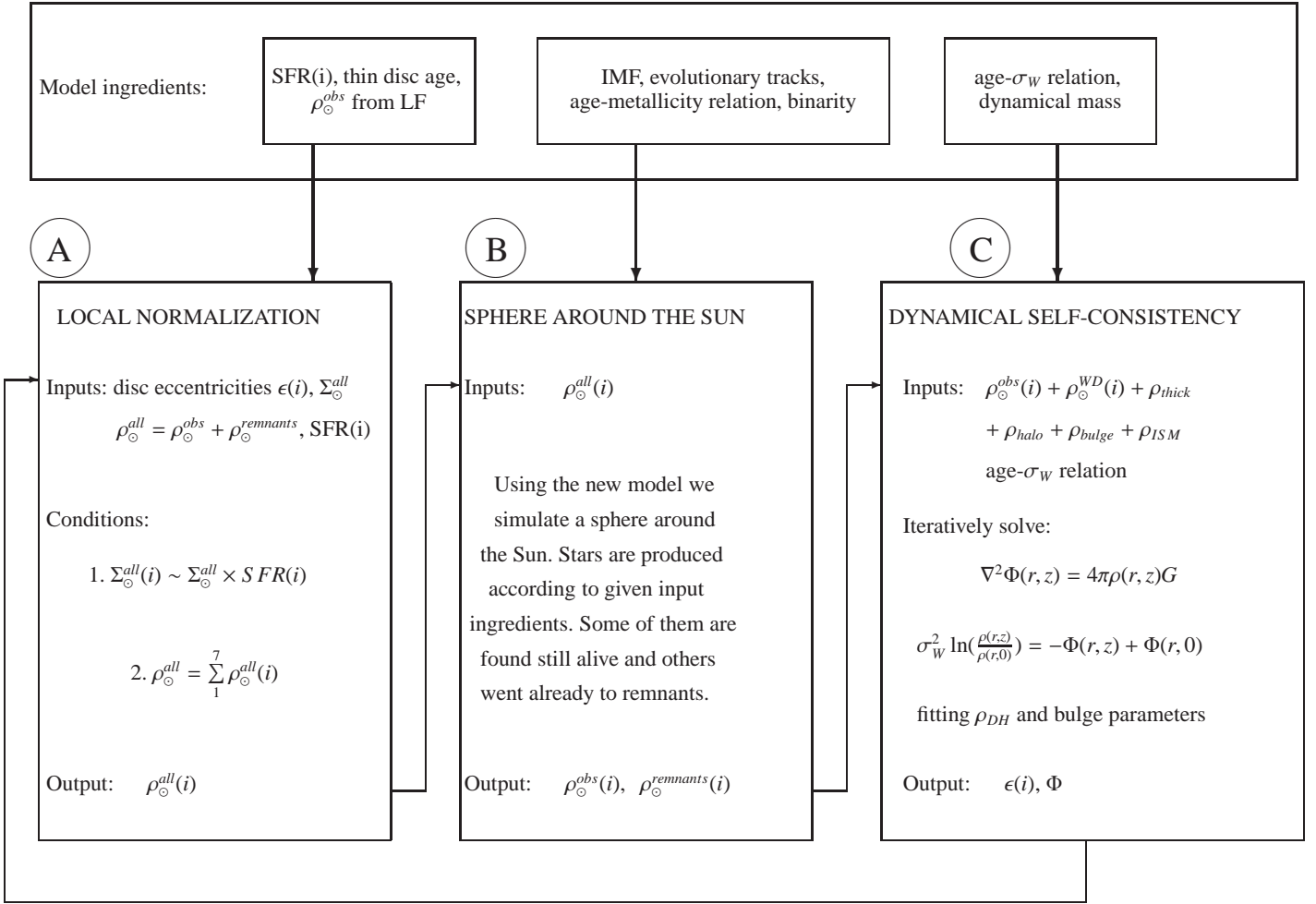


The process 
of determining the mass model parameters is presented in Fig. \ref{massModel}. Block A is a local mass 
normalization, which provides the values of local volume mass density $\rho_{\odot}^{all}(i)$ 
for each age subcomponent $i$. Block B is where we perform the simulations of the local neighborhood and 
derive the percentage of alive stars and remnants for each thin disc subcomponent. Thus, the procedure presented 
in blocks A and B simply splits the stellar local volume mass density $\rho_{\odot}^{obs}$ 
(at all ages) into seven disc subpopulations according to the SFR and includes the secular heating process 
 to get $\rho_{\odot}^{obs}(i)$. In Block C, we impose the dynamical constraints 
to get the excentricities of the ellipsoidal density distribution $\epsilon(i)$ from the Boltzmann equation. The whole procedure is a 
normalization iterative process, which uses the observational and dynamical constraints, provides the values of 
$\rho_{\odot}^{obs}(i)$ and $\epsilon(i)$, and that constructs the mass model to be applied in simulations in this way.
Initially, some estimates are assigned to eccentricities $\epsilon(i)$ and to the local surface density
$\Sigma_{\odot}^{all}$. Due to the subdivision of the thin disc into seven age subcomponents, the SFR is introduced 
into the model by the means of the intensity of star formation history at each epoch SFR(i). Block A requires two 
conditions to be fulfilled. The first one is that the surface density of each age subcomponent 
$\Sigma_{\odot}^{all}(i)$ has to be proportional
to the intensity of SFR in its corresponding age bin SFR(i), and the second one requires that the sum of calculated 
volume densities of all disc subcomponents $\rho_{\odot}^{all}(i)$
must fit the value of the imposed total volume density $\rho_{\odot}^{all}$. When the iteration 
finishes we get the local volume density split into seven components  $\rho_{\odot}^{all}(i)$. Subsequently, we 
perform the simulations of a sphere around the Sun to get the values of the local volume
 density of alive 
stars split into seven components $\rho_{\odot}^{obs}(i)$. In Block C, we treat the dynamics: this is where the 
disc eccentricities $\epsilon(i)$ are calculated and the dynamical self-consistency is ensured. The full description 
of this process is given in \cite{bienayme}. The values of the velocity dispersion  $\sigma_W$ for each subcomponent 
$i$ are adopted from observations (see Section \ref{agemet}). First, all mass components-thin disc $\rho_{\odot}^{obs}(i)$, thin disc white dwarfs $\rho_{\odot}^{WD}(i)$, thick disc $\rho_{thick}$,
halo $\rho_{halo}$, bulge $\rho_{bulge}$ and interstellar medium $\rho_{ISM}$-enter the 
Poisson equation and impose a potential $\Phi$. Then the dark halo ($\rho_{DH}$) and bulge parameters are
adjusted until the potential produces an acceptable rotation curve. The new potential combined with the velocity 
dispersions in the Boltzmann equation produce new values of $\epsilon(i)$ modifying the density distribution
of each stellar component. The process is iterated until the mass model stabilizes by producing a self-consistent 
solution. It is important to emphasize that each time an evolutionary 
ingredient of the model changes (see Fig. \ref{ingredients}) the mass model has to be rederived.

\subsection{The code and its implementation}\label{sec:CodeOrganization}

\subsubsection{The thin disc treatment}\label{imf-des}

\begin{figure} 
\begin{center}
\includegraphics[width=9cm]{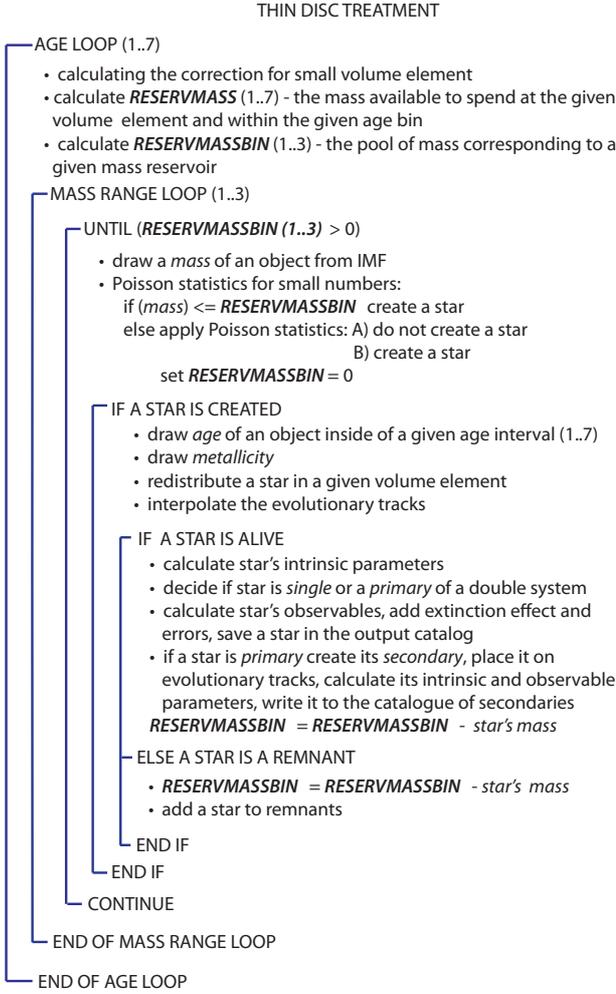}\\
\end{center}
\caption{The organigram of the new model's thin disc treatment. }
\label{organigram2}
\end{figure}

The BGM provides star-counts prediction towards a selected direction in the sky. Either a line of sight or a 
zone can be simulated. The outermost loops of the new code have not changed, and as in \cite{ar2003}, they are the
coordinates and distance loops. Together, they form the volume element loop; thus, stars are produced with distance 
steps while moving from the Sun until the chosen distance limit. In each volume element, the objects belonging to 
the basic Galactic components are generated. Here, we present and discuss the new thin disc treatment.  We start with the general outline of the new code, which is sketched in Fig. \ref{organigram2}. 

The outermost loop of the thin disc treatment is the age loop. Each disc's subcomponent -- seven for the time being -- 
simulates an age interval to cover the total disc age in fine resolution. In each age bin, the mass available to be spent on star production in a given volume 
element is the \textit{mass reservoir}. It is the amount of mass that is expected to be found in 
stars at the specified position (volume element) according to the predefined evolutionary
 (IMF and SFR) and density 
(density laws) parameters. It is calculated from the expression 
\begin{equation} mass ~reservoir = dV \times \rho(x,y,z,i),
\label{reservmass1}
\end{equation} 
where $i$ denotes the thin disc age subcomponent. This equation is the backbone of the thin disc simulation.
The densities $\rho(x,y,z,i)$ are obtained taking into account the density law, SFR, and secular heating that corresponds 
to a given 
$i$-th subcomponent after performing the local mass normalizations. Then, the \textit{mass reservoir}, which is calculated for 
a given volume element and disc subcomponent, is divided into three bins corresponding to three 
IMF ranges. By construction 
the model works with three mass ranges, and it allows the introduction of at most three-slope {\it power law} IMFs.
 This division is directly related to the power law description of the IMF,which is most commonly used in literature,
 and has proved to fit the observational data (\cite{Kroupa1993}).  In the future, other 
 analytical functions, such as the ones proposed by \cite{Chabrier2005}, can be easily implemented as well.
The drawing of the stars is done separately for each of the three mass intervals always by starting 
 from the
high mass bin (see the mass range loop in Fig. \ref{organigram2}). To divide the mass
 reservoir into three parts that
correspond to three IMF slopes, mass normalization factors are computed. They are 
presented later in this section.

At short distances from the Sun, the volume elements for a given line of sight are small. Consequently, the mass available to spend 
on stars is also small and could introduce a bias towards creating too many low-mass objects. Basically, whenever the 
mass is not sufficient to produce a more massive star, too many low mass ones are drawn instead, thus  
significantly biasing the shape of the IMF. To avoid that bias, we have implemented a correction for the small 
volume elements, as explained in Appendix \ref{volume}.

In each mass range, the star production proceeds until there is no more mass to convert into stars 
(null mass reservoir). We verify that there is no bias in the masses drawn in each mass range. Once the mass of an 
object is drawn from the given IMF, we check if the reservoir of mass is enough
to create this object. The high mass bin corresponds to the long and steep tail of the IMF, so it can happen 
that a mass drawn for a star candidate exceeds the total mass available in that case, although resources are relatively big. 
A similar situation applies to the remaining two mass regions, but with a smaller rate of occurrence. Losing that mass would 
produce a bias in the mass distribution. To correct for that, we introduce a random drawing, according to Poisson 
statistics, to either generate or not 
this drawn star in the remaining \textit{mass reservoir}, and we correct the reservoir value accordingly. 
 We verify a posteriori that the drawn masses and the IMF are correct globally.

Once an object is created we assign to it an age and metallicity. The ages are drawn randomly from the uniform distribution
in the interval of the given age subcomponent. The metallicity is drawn for each star from its own age, according to 
the age-metallicity relation adopted, while the mean age of the bin was assigned instead in the previous version of BGM. 
This results in a smoother age-metallicity relation in the resulting simulated stars. 

When the age, mass, and metallicity 
are established, we interpolate the evolutionary tracks and find the position of the star in the HR diagram. If the
solution is found, we consider that a star is alive, while the star is added to the remnant
pool if a solution is not found. If the star is alive, its intrinsic parameters are calculated and then, according to a given probability, it is 
decided if that star is single or if it is the primary component of a binary system. Subsequently, the observables are 
assigned to that object and the extinction effect is included. Optionally, the observational errors are added, and the star 
is written to the catalogue. If the 
produced star was flagged as a primary, we create its secondary, place it on the evolutionary tracks, calculate its 
intrinsic parameters, and write it to the catalogue of secondaries. The mass of the second component is subtracted
from the $mass ~reservoir$. The merging of binary systems is performed later (see Section \ref{subsec:binarityImplementation}). We then keep producing 
stars subsequently from the three mass ranges as long as there is mass left in the given pool. The
procedure continues for all age subcomponents and for all volume elements.

The IMF takes the form of three segments of power law, where for each one,

\begin{equation}
\phi(m) = m^{-(1+x)},
\label{im}
\end{equation}
where $m$ is the mass, $x$ is the slope and $(1+x)$ is also commonly denoted as the $\alpha$ parameter.
Our code is designed to deal with three slope initial mass function. To get the mass 
locked within each mass interval one must solve $\int m \: \phi(m) dm$ for all three ranges. The sum of those three 
numbers must be normalized to one. The relative mass locked within each mass interval, $M_{Int_i}$  is computed as follows:
\begin{equation}
M_{Int_i} = K_i \int_{m_i}^{m_{i+1}} m \phi(m) dm,
\label{fin}
\end{equation}
where $K_i$ are the continuity coefficients. The sum of those values is normalized to one, so when we multiply the 
\textit{mass reservoir} (total mass within a given volume element) by each of them, we distribute the available mass over 
three ranges in a given volume element for a given subpopulation.

\subsubsection{Binarity}\label{subsec:binarityImplementation}

The new BGM is able to generate binary systems. Binaries are implemented by following 
the scheme proposed by \cite{Arenou2011}, which is applied in the Gaia simulator (\cite{RobinLuri2012}).
The 
thorough explanation of the algorithm and the observational data, which were used to set the empirical relations
adopted in this scheme, can be found in those articles. 
Here, we provide the general description of how that procedure was implemented in the model. It is presented in the 
core of the thin disc treatment in Fig. \ref{organigram2}.
Every new created star is decided to be a single or a primary component of a double system according to a probability, 
which depends 
on the object's mass and luminosity class (see Section \ref{sec:binarityDa}).
If a star, which 
has just been created is determined to be a primary, we subsequently create its secondary using the following steps. 
The separation of the system is estimated from the probability distribution function of the semi-major axis  
derived empirically from observations (see \cite{Arenou2011}). The mass ratio statistics are also
estimated from observations, which takes into account the spectral type of the primary and the binary period. 
 We want to emphasize that we have preferred to apply the empirical law for the derivation of the
mass distribution of the secondary stars instead of a theoretical imposed IMF. The small differences between these two
approaches are evaluated in Appendix \ref{IMFsec}.
We assign the same age and metallicity to the secondary component as the
primary object, and then it is placed on the evolutionary tracks and its 
intrinsic parameters are determined. We are not considering interacting binaries, 
which may follow different evolutionary tracks due to the presence of the companion. 
The secondary star is always found alive, since it has been already checked at this point 
that its primary, which is more massive, is alive.  

After its creation, the system is randomly inclined 
with respect to the line of sight and then is projected on the sky. Subsequently, the angular 
separation is computed and the decision of whether the system is going to be a resolved or unresolved binary system 
depends on the imposed resolution of the catalogue.

It must be emphasized that the constraint on the local 
stellar mass density is conserved when generating the binaries. It is a strong advantage of the new BGM scheme that the 
total mass is constrained previously to simulations 
through the dynamical self-consistency calculations. This means that the mass is established before and after the
stars, whether single or double, which are created from the available \textit{mass reservoir}.

\subsection{New processing modes}\label{sec:modes}

The BGM has several different modes in which simulations can be performed. There was a 
strong need 
for performing volume limited simulations in the solar neighbourhood, especially when reproducing the local observed 
LF  and calculating the SFR normalization in a cylinder. For that, we have incorporated two new
 processing modes 
in the model: we can simulate a sphere (with a constant spatial density) or a cylinder centred at the Sun position
and perpendicular to the Galactic plane. 
The spherical volume element is used to calculate the synthetic LF and compare it to the observed one. Using 
a cylinder mode, we have tested the algorithms of star production and checked if we are able to reproduce the imposed
theoretical SFR integrated in the perpendicular direction $z$.

\section{Observed data sets}\label{tools}\label{chapter:fitting}

We decided to compare simulations by two observational sets: the Tycho-2 data up to 
the magnitude $V_T$ = 11.0
and the single star luminosity function LF. 
In the first case, we were interested to compare the simulated star counts and the $(B-V)_T$ distributions 
with Tycho-2 data over the whole sky.
In the second case, we have performed the simulations of the local sphere 
(see Section \ref{sec:modes}) for each model and obtained the synthetic LF, which were compared with the observed LF.

\subsection{The Tycho-2 Catalogue}
\label{tyc}
The Tycho-2 catalogue \citep{tycho2} provides positions, 
proper motions, and $B_T$ and $V_T$
 magnitudes of the 2.5 million 
brightest stars across the entire sky. Stars until magnitude $V_T$ $\sim$ 11.0 were chosen ensuring a completeness of  99 
percent up to this magnitude (864,816 stars). We decided to work in the space of observables, so
 we transformed our 
simulations done in the Johnson photometric system into the Tycho system. Different photometry 
transformations were 
evaluated (see Appendix \ref{photo}). 

As it is shown in Section \ref{res}, Tycho data shows a bi-modal colour distribution. In the blue 
peak, we mainly have B, A, and F type main-sequence stars, which are on average young, while the red peak is dominated by 
 giant stars covering  a wide age range. Most of the stars fall within the absolute magnitude range -1 $< M_V <$ 5  mag.

To reduce the computational time, the systematic comparison between simulated and Tycho-2 data was done by 
considering three established reference regions: 
the whole Galactic north pole at  b $> 70{^\circ}$; the region at intermediate latitude at 160${^\circ}$ $<$ l $<$ 180${^\circ}$ 
and 20${^\circ}$ $<$ b $<$ 40${^\circ}$ and the region within the plane at 70${^\circ}$ $<$ l $<$ 90${^\circ}$ and 
-10${^\circ}$ $<$ b $<$10${^\circ}$. These regions have been selected to be representative of
 the full longitude range at the corresponding latitude, meaning that their colour distribution shows the same
 characteristics as one of the full longitude stripes at the specified latitude. In addition to these regions, whenever a
 more general comparison 
was required,  a whole-sky analysis was performed and eventually,  sky maps of mean colours and relative number 
of objects was produced.

\subsection{The local luminosity function}

Three observed LFs have been considered: \cite{Jahreiss1997},
\cite{Kroupa2001}, and \cite{Reid2002}. All of them are the LF of single stars. 
Whereas \cite{Kroupa2001} and \cite{Reid2002} provide only the LF 
for the stars on or near the main sequence, \cite{Jahreiss1997} also provides the total LF, which included the evolved stars. 
Differences among them are observed at both, the bright end and faint end of the LF.

As we work with Tycho data, we are interested in the bright side of the LF, that is, at $M_V <$ 5. 
In this region, 
although the Hipparcos survey provided completeness at a 25 pc distance limit, 
  one can notice some differences in LFs from different authors. These differences reflect Poisson uncertainties 
  and are mainly attributed to the slightly different volume considered by each author.


\section{Model ingredients}
\label{sec:ingred}

In this section, we describe the analytical expressions and the empirical relations tested for all the model 
ingredients detailed in Table \ref{tableParam}.
As indicated in Fig. \ref{massModel}, a new mass model was derived and the Galactic gravitational potential was recalculated when any of the model's ingredients
listed in the top boxes of Fig. \ref{massModel} were changed. 

\subsection{The initial mass function and star formation history}\label{sec:imfsfrRes}

Table \ref{tabIMF} lists the eleven IMFs, which were tested in our study. All the considered IMFs are power-laws  
(see Eq. \ref{im}), which differ 
only in the number and values of slopes and the corresponding mass ranges.

\begin{table*}
\begin{center}
\begin{tabular}{|c|c|c|c|c|c|c|c|}
\hline \hline
\multicolumn{1}{|c|}{IMF}&\multicolumn{1}{|c|}{M$_1$}&\multicolumn{1}{|c|}{$\alpha_1$}&\multicolumn{1}{|c|}{M$_2$}&\multicolumn{1}{|c|}{$\alpha_2$}&\multicolumn{1}{|c|}{M$_3$}&\multicolumn{1}{|c|}{$\alpha_3$}&\multicolumn{1}{|c|}{M$_4$} \\
\hline
 \cite{Scalo1986}     &  0.09  & 1.25  & 1.0   &  2.35  & 2.0  & 3.0  & 120 \\
 \cite{misha1997-1}   &  0.09  & 1.7   & 1.0   &  2.5   & 3.0  & 3.0  & 120\\
 Haywood-Robin        &  0.09  & 1.6   & 1.0   &  3.0   & -    & 3.0  & 120\\
 \cite{Vallenari2006} &  0.09  & 1.1   & 0.8   &  2.3   & -    & 2.3  & 120\\
 \cite{Kroupa2008}    &  0.09  & 1.3   & 0.5   &  2.3   & -    & 2.3  & 120\\
 \cite{JJ2010}        &  0.09  & 1.46  & 1.72  &  4.16  & -    & 4.16 & 120 \\
 KH-v1                &  0.09  & 1.3   & 0.5   &  2.3   & 1.53 & 3.0  & 120\\
 KH-v4                &  0.09  & 1.3   & 0.5   &  2.3   & 1.53 & 3.5  & 120\\
 KH-v6                &  0.09  & 1.3   & 0.5   &  1.8   & 1.53 & 3.2  & 120\\
 KH-v7                &  0.09  & 1.3   & 0.8   &  2.3   & 1.53 & 3.0  & 120\\
 KH-v8                &  0.09  & 1.1   & 0.8   &  2.1   & 1.53 & 3.0  & 120\\
\hline
\end{tabular}
\end{center}
\caption{ The list of eleven IMFs applied in our simulations. The M$_1$, M$_2$, M$_3$ and M$_4$ are the four limiting masses, expressed 
          in $M_\odot$ and  $\alpha_1$, $\alpha_2$ and $\alpha_3$ are the corresponding slopes. 
          The values of M$_1$ 
          and M$_4$ were fixed according to the limiting masses of the evolutionary tracks. }
\label{tabIMF}
\end{table*}

From literature, we considered the following six IMFs: \cite{Scalo1986}, \cite{misha1997-1}, 
\cite{misha1997-2} with a corrected slope at the
 low masses as proposed by \cite{ar2003} (hereafter called Haywood-Robin IMF), \cite{Vallenari2006}, \cite{Kroupa2008}, and 
\cite{JJ2010}. We extended the low mass range of \cite{Vallenari2006} IMF  from 0.2 to 0.09 $M_\odot$.
After several tests, we proposed five new IMFs, which are different combinations of  \cite{Kroupa2008} at low masses 
and \cite{misha1997-2} at high masses. To merge the two functions, we looked for 
their intersection above 1 $M_\odot$. It was found to be at 1.53 $M_\odot$.
The IMF called KH-v1 (from Kroupa-Haywood version 1) keeps the IMF of \cite{Kroupa2008} below that value and  
the \cite{misha1997-2} above. Subsequently, other IMFs were defined by slightly changing the slopes and/or the 
limiting masses. Table \ref{tabIMF} presents the five that provide a better fit to Tycho data.  

\begin{figure}
\begin{center}
\includegraphics[width=5cm, angle=-90]{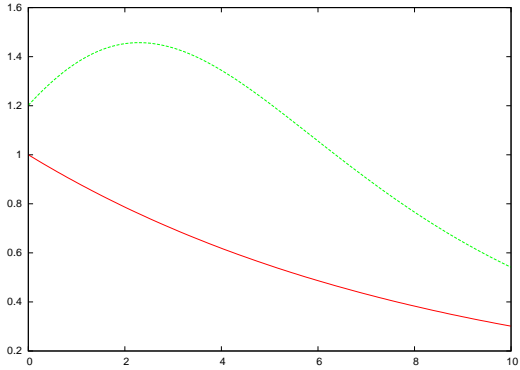}\\
\end{center}
\caption{ Two decreasing SFRs as a function of time $t$. The SFR of model A by
          \cite{JJ2010}  (green dashed line) and the (\cite{AumerBinney2009}) single exponential with  $\gamma=0.12$ 
          (red solid line). Normalizations are arbitrary.}
\label{sfrs}
\end{figure}

Constant and decreasing SFRs are considered. To test a declining scenario, we used two 
approaches. 
We considered the \cite{AumerBinney2009} simple exponential function SFR$(\tau) \propto \exp (-\gamma t)$, 
where $\gamma$ is a parameter and $t$ is time with $\gamma$ to be 0.12 
(one of the \cite{AumerBinney2009} good fits). As a second approach, we took the  more complex expression proposed 
by  \cite{JJ2010}. Those two decreasing SFRs are depicted in Fig. \ref{sfrs}.

\subsection{Evolutionary tracks sets}

Table  \ref{tabTracks} lists the different sets implemented in the BGM since 1986.
As mentioned in Fig. \ref{massModel},  we tested two evolutionary tracks sets named package E1 and E2 (see Table  \ref{tabTracks}).  
Package E2 uses \cite{Bertelli2008} tracks for masses
 $\geq 0.7 M_\odot$, and \cite{Chabrier1997} tracks at lower masses.

\begin{table}[ht!]
\begin{center}
\begin{tabular}{|c|c|}
\hline \hline 
 Model version & Evolutionary tracks     \\
\hline \hline
                      & \cite{Mengel1979} for $M < 3 ~M_\odot$; \\
  \cite{ar1986} & \cite{Sweigart1978} for the giants; \\
                      & \cite{Becker1977} for $ 3 - 15 ~M_\odot$   \\
  & and \cite{Chiosi1978} for $M > 15~ M_\odot$; \\
\hline
  & \cite{Schaller1992} for stars $ M > 1 ~M_\odot $;\\
 \cite{misha1997-1}  & Vandenberg smaller masses;\\
  \cite{ar2003}                   & \cite{Castellani1992} for \\
  & helium-burning stars at masses $1 - 1.7 ~ M_\odot $; \\
\hline
          & Chabrier and Baraffe models previous  \\
  & to publication in 1997 for $ M < 0.6 ~ M_\odot $; \\
   package E1                 &  \cite{Bertelli1994} \\
  & for $ 0.6 ~ M_\odot <  M < 120.0 ~ M_\odot$\\
\hline
           & \cite{Chabrier1997} models \\
  & for low-mass stars with $  M < 0.7 ~ M_\odot $;\\
       package E2          & \citet{Bertelli2008, Bertelli2009} for high\\
 &  masses with $ 0.7 ~ M_\odot < M < 20~M_\odot  $ ;\\
    &    \cite{Bertelli1994}  \\ 
    &   for $ 20~  M_\odot <  M < 120.0 ~ M_\odot;  $           \\ 
\hline
\end{tabular}
\end{center}
\caption{Evolutionary tracks used in subsequent versions of BGM.}
\label{tabTracks}
\end{table}

\subsection{Binarity}\label{sec:binarityDa}

To compare our simulations with Tycho-2, data a spatial resolution of  0.8 arcsec was assumed,  meaning that all the binary systems generated below this angular separation
were considered as unresolved. We have checked that values of 0.5 or 1.0 arcsec do not change much
colour histograms as  most of the simulated binaries turn out to have much smaller separations.

\subsection{The local stellar and ISM mass densities}
\label{sec:LMD}

In our model, the local stellar mass density of the thin disc component is a critical parameter 
(see Fig. \ref{massModel}). In Table \ref{rhoObs1},  we present the most used values derived  
from observations:  \cite{Wielen1974}, hereafter referred as LMD1, and \cite{Jahreiss1997}, 
which are obtained using Hipparcos data and are referred here as LMD2. As can be seen in Table \ref{rhoObs1},
 this last value is in good agreement with the value derived by \cite{Reid2002}, which is also from Hipparcos data. 
We varied only the density corresponding to main sequence and giant stars ($\rho_{\odot}^{obs}$), 
which maintains a value 0.007 $ M_\odot pc^{-3}$ for white dwarfs in both cases. Thus, the tested values for $\rho_{\odot}^{obs}$ 
are 0.039 $M_\odot pc^{-3}$ (LMD1) and  0.033 $M_\odot pc^{-3} $ (LMD2).
The way in which the local stellar mass density is distributed over the seven subcomponents of the thin disc 
is illustrated in Fig.  \ref{massModel}.

In \cite{ar2003}, the mass density 
assumed for the interstellar medium was $0.02 M_\odot pc^{-3}$. In the present work, we change that value to 
$0.05 M_\odot pc^{-3}$ as proposed by \cite{BT2008}. As it is known, this parameter is very uncertain and its 
influence in the process of the derivation of the total dynamical mass is discussed in Section \ref{dynmass}.

\begin{table}
\begin{tiny}
\begin{center}
\begin{tabular}{|c|c|c|c|}
\hline \hline 
\multicolumn{1}{|c|}{Source}&\multicolumn{1}{|c|}{\cite{Wielen1974}}&\multicolumn{1}{|c|}{\cite{Jahreiss1997}}&\multicolumn{1}{|c|}{\cite{Reid2002}}\\
\multicolumn{1}{|c|}{}&\multicolumn{1}{|c|}{LMD1}&\multicolumn{1}{|c|}{LMD2}&\multicolumn{1}{|c|}{}\\
\hline \hline 
ms & 0.038  & 0.0323  &  0.0300-0.0338 \\
giants & 0.001  & 0.0006  &    \\
WD & 0.007  & 0.0053  & 0.004    \\
total & 0.046  &  0.039  &     \\
\hline
\end{tabular}
\end{center}
\caption{ The values of the thin disc local stellar volume mass density derived from observations (Units:  $M_\odot pc^{-3}$). }
\label{rhoObs1}
\end{tiny}
\end{table}

\subsection{Atmosphere models}
Three different sets of atmosphere models were tested: the BaSeL 2.2 
\citep{Lejeune1997,Lejeune1998}, BaSeL 3.1 \citep{Westera2002}, and the giant MARCS grid \citep{Houdashelt2000}. In the last case, MARCS models for giants were combined with BaSeL3.1 for other stars.

\subsection{Age-metallicity and age-velocity relations}\label{agemet}

Two age-metallicity relations have been considered, the one from \cite{Twarog1980} (used in the 
old model) and the one
proposed by \cite{Haywood2006}. Twarog's relation assigns significantly lower metallicities 
than Haywood's one.
Straight lines were fitted to both relations and dispersion curves. We checked that 
the new mechanism (see Section \ref{imf-des}) of assigning the metallicity diminishes the 
step-like form 
and smooths out the resulting relation. 
The intrinsic scatter of the metallicity is taken into account. The scatter has a specific value for each age and follows 
the relation given by \cite{Haywood2006} in his Figure 13c. 
This scatter is increasing with age, a trend that has been interpreted
as due to the pollution by stars that come from both the inner and outer
disc (\cite{Haywood2008}; \cite{Haywood2013}).

Table \ref{tableSigmaW} shows the values of the two age-velocity relation (AVR)
tested here, the \cite{gomez} relation, used in the old model, and the new \cite{Holmberg2009} relation, 
which  have significantly higher velocity dispersion perpendicular to the plane, for the old disc. 
\begin{table} [h!]
\begin{tiny}
\begin{center}
\begin{tabular}{|c|c|c|c|}
\hline \hline 
\multicolumn{1}{|c|}{C }&\multicolumn{1}{|c|}{Age}&\multicolumn{1}{|c|}{\cite{gomez} }&\multicolumn{1}{|c|}{ \cite{Holmberg2009} }\\
\multicolumn{1}{|c|}{}&\multicolumn{1}{|c|}{[Gyr]}&\multicolumn{1}{|c|}{$\mathbf{\sigma_W}$ [km/s]}&\multicolumn{1}{|c|}{ $\mathbf{\sigma_W}$ [km/s] }\\
\hline \hline 
 1  & 0-0.15  & 6.0   & 6.0    \\
 2  & 0.15-1  & 8.0   & 8.0     \\
 3  & 1-2     & 10.0  & 10.0    \\
 4  & 2-3     & 13.2  & 14.0     \\
 5  & 3-5     & 15.8  & 17.5    \\
 6  & 5-7     & 17.4  & 21.0     \\
 7  & 7-10    & 17.5  & 25.0    \\
\hline
\end{tabular}
\end{center}
\caption{ The comparison of the age-velocity dispersion relation
by \cite{gomez} and \cite{Holmberg2009}. C stands for the thin disc subcomponent. }
\label{tableSigmaW}
\end{tiny}
\end{table}

\subsection{The age of the thin disc}

In the old BGM, the age of the formation of the thin disc was set to 10 Gyr. Two more values, the values of 12 Gyr 
(one of the favoured values by \cite{AumerBinney2009})
and 9 Gyr, proposed by  \cite{delPeloso}, who derived an age of 8.8 $\pm$ 1.7 Gyr from Th/Eu nucleocosmochronology
were tested here.

\subsection{The thick disc}\label{sec:thick}

As described in \cite{Reyle2001}, the thick disc density law is assumed to be a truncated exponential characterised by three
parameters: $h_z$ the scale height, $\rho_{thick}$ the local density and $x_l$ the distance above the plane, where the density law 
becomes exponential.  
\cite{Reyle2001} have shown that there is a degeneracy in the parameter determination because the local 
density and the scale height are anti-correlated when using star counts at high latitudes as constraints. 
Hence, they consider a family of solutions having a fixed parameter df $\propto$ local density 
$\times ~ h_z^{2}$.
We have tested two solutions here: model TkD1 with 
$h_z$=1200 pc, $x_l$=72 pc, and  $7.23 10^{-4} ~ M_\odot pc^{-3}$ and model TkD2 with 
$h_z$=800 pc, $x_l$=400 pc, and $\rho_{0}= 2.9 10^{-3} ~ M_\odot pc^{-3}$.
 The $x_l$ parameter is fitted using the Boltzmann equation.
The metallicity assumed for the thick disc is -0.48 dex and the dispersion is 0.30.
 The comparison of the two presented models is shown in Section \ref{add} and discussed in Section \ref{summary}.

\subsection{Extinction model}\label{sec:extinction}

Both the \cite{drimmel2001} and the \cite{marshall3D2006} extinction models are implemented in the BGM. 
\cite{marshall3D2006} covers only the region  
-100${^\circ}$ $<$ l $<$ 100${^\circ}$  and -10${^\circ}$ $<$ b $<$ 10${^\circ}$,
whereas the \cite{drimmel2001} extinction model covers the full sky and has saturation problems in the complex regions 
near the Galactic centre. 

\subsection{Radial scale length}
\label{radsl}

In the old model, the values of radial scale length were fitted to  $h_R$ = 5000 pc for the young disc subpopulation 
($\tau <$ 150 Myr) and $h_R$ = 2400 pc for older stars. We shall refer to 
these values as model SL1.
To check the influence of this parameter, alternative values were tested. A large scale length for the young population 
was implemented:  $h_R$ = 5500 pc for the stars with $ \tau <$ 150 Myr; $h_R$ = 5000 pc for stars with 
150 Myr $< \tau <$ 1 Gyr; $h_R$ = 3500 pc for 1 Gyr $< \tau <$ 2 Gyr and $h_R$ = 2400 pc for $\tau >$ 2 Gyr 
(model SL2). The aim was not to adjust the scale length of the thin disc, which would be difficult with Tycho data only, but to test its influence in the present analysis on the star counts and colour distribution within the Galactic plane.

\section{Results}\label{res}

\subsection{The old model vs. Tycho-2}

\begin{figure} 
\begin{center}
\includegraphics[width=9cm]{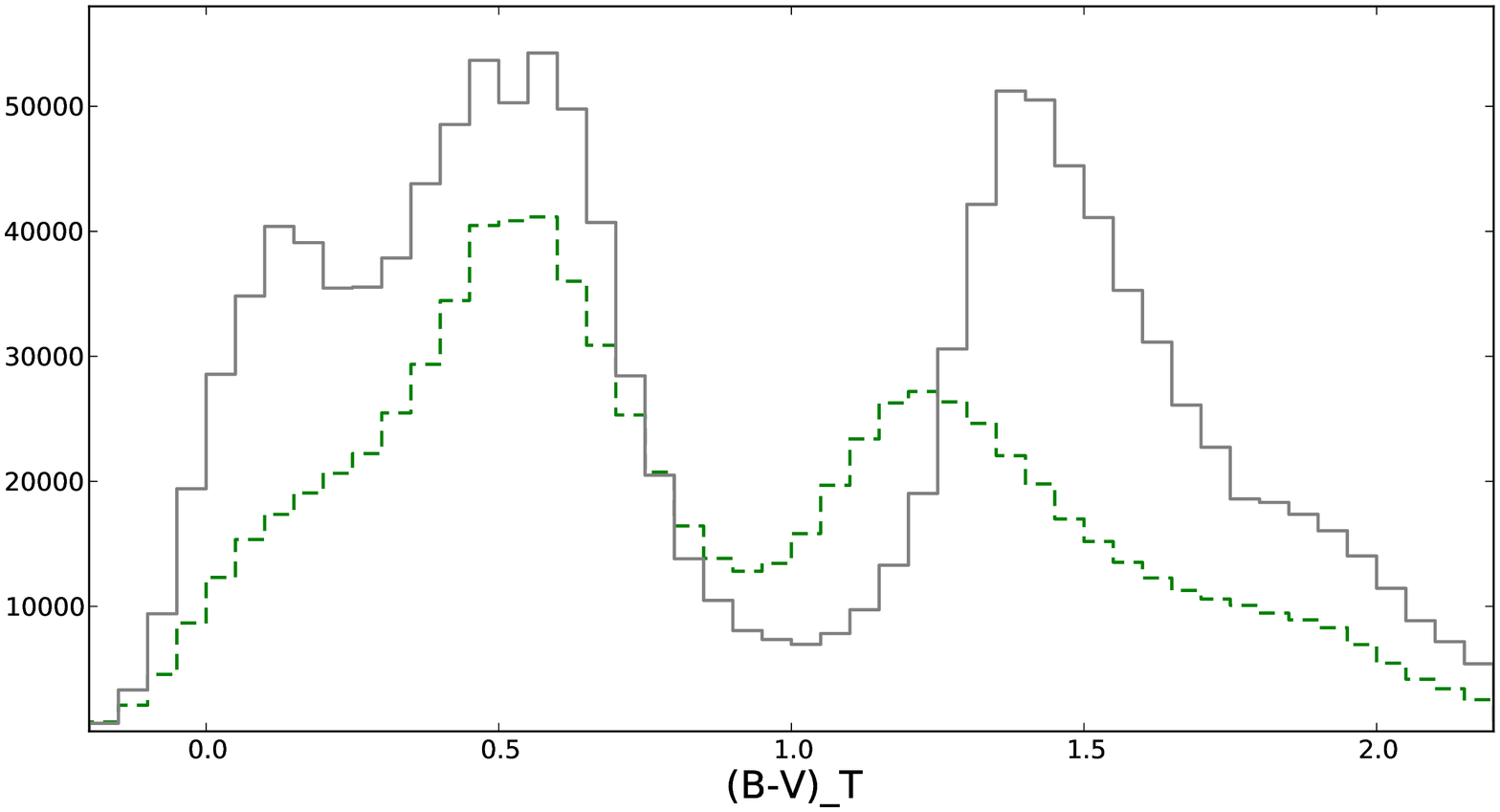}\\
\end{center}
\caption{Whole sky $(B-V)_T$ distributions of the old model (grey solid line) and Tycho-2 data (green dashed line).}
\label{oldModel}
\end{figure}

Figure \ref{oldModel} presents the comparison between the colour distributions from the old BGM from Robin et al. (2003) 
and from Tycho-2 data. The ingredients of the old model are presented synthetically in the second column of 
Table \ref{tableParam}. When comparing the old model to Tycho-2 data, one notices several 
important discrepancies:

\begin{enumerate}
 \item The red peak of the modelled $(B-V)_T$ distribution is shifted by about 0.2 mag to the red. 
 \item The excess of the total number of stars: the model produces two times more stars than the Tycho-2 sample.
 \item The excess of the blue stars around $(B-V)_T \sim$ 0.15 mag.
\end{enumerate}

\begin{table*}
\begin{center}
\begin{tabular}{|c|c|c|c|}
\hline \hline \hline
\multicolumn{1}{|c|}{Ingredients}&\multicolumn{1}{|c|}{Old model}&\multicolumn{2}{|c|}{New default models}\\
\multicolumn{1}{|c|}{}&\multicolumn{1}{|c|}{}&\multicolumn{1}{|c|}{model A}&\multicolumn{1}{|c|}{model B}\\
\hline
 IMF  &  Haywood-Robin & Haywood-Robin (\textit{A})  &  Kroupa-Haywood v6 (\textit{B}) \\
\hline
 SFR  & constant  & a decreasing $\exp (-0.12 \tau)$  & a decreasing $\exp (-0.12 \tau)$       \\
      &  & \cite{AumerBinney2009}  & \cite{AumerBinney2009}      \\
\hline
 evolutionary   & see Table \ref{tabTracks}   & package E2 Table  \ref{tabTracks} & package E2 Table  \ref{tabTracks}      \\
  tracks    &   &      & \\ 
\hline
 age-metallicity   & \cite{Twarog1980}  & \cite{Haywood2006} & \cite{Haywood2006}      \\
 relation     &   &    &  \\
\hline
 atmosphere   &  BaSeL 2.2  & BaSeL 3.1   & BaSeL 3.1      \\
    models  &   &    &   \\
\hline
 binarity  & no  & yes: from \cite{Arenou2011} & yes: from \cite{Arenou2011}     \\
\hline
 thin disc age &  10 Gyr & 10 Gyr & 10 Gyr       \\
\hline
 thick disc  &  $x_l$ = 400 pc, $h_z$ = 800 pc &  $x_l$ = 400 pc, $h_z$ = 800 pc &  $x_l$ = 400 pc, $h_z$ = 800 pc \\
  parameters    & density = 0.0083 */$pc^{3}$ & density = 0.0083 */$pc^{3}$ & density = 0.0083 */$pc^{3}$  \\
\hline
 extinction   & \cite{drimmel2001} & \cite{drimmel2001} & \cite{drimmel2001}    \\
  model    &   & + \cite{marshall3D2006} & + \cite{marshall3D2006}      \\
\hline
 ISM local   & \cite{ar2003} &  \cite{BT2008} & \cite{BT2008}   \\
    density  &   &    &   \\
\hline
 local stellar   &  \cite{Wielen1974} & \cite{Wielen1974}  & \cite{Jahreiss1997}    \\
   mass density   &                   &     & \\
\hline
 age-velocity   &  \cite{gomez} &  \cite{gomez} &  \cite{gomez}       \\
  relation    &   & &        \\
\hline
 warp   & \cite{Reyle2009} &  \cite{Reyle2009} & \cite{Reyle2009}       \\
\hline
 scale length   & young disc $h_R$ = 5000.0 pc & young disc $h_R$ = 5000.0 pc & young disc $h_R$ = 5000.0 pc    \\
      & old disc $h_R$ = 2400.0 pc &  old disc $h_R$ = 2400.0 pc & old disc $h_R$ = 2400.0 pc     \\
\hline
\end{tabular}
\end{center}
\caption{Most important ingredients of the old and new model, which have been analysed. 
         Notice that the two default models differ only by the IMF and the value of the local 
stellar mass density.
         ($^1$) The \cite{drimmel2001} extinction model was not used in the BGM from \cite{ar2003}, but it was used
         here in the old model simulations. }
\label{tableParam}
\end{table*}

\subsection{Defining a default model}
\label{chi}

Our strategy was to analyse the impact on the results when each of the ingredients presented in Section 
\ref{sec:ingred} was changed 
independently from the others. We must emphasize that several ingredients are 
highly correlated. Thus, we also study the combinations of parameters which are correlated, for example, 
 the IMF, SFR and the local mass density.
 A detailed analysis, simulations of different regions, and comparisons with Tycho-2 data allowed us to establish and 
justify the composition of a {\it default model}, defined as a combination of a new set of ingredients that significantly 
improve the fit to Tycho data. 
 In the third and fourth columns of Table \ref{tableParam}, we present the list of these ingredients. 
We present two default models, A and B because there are two IMFs that we favour, as shown later on
in Section \ref{masstime}. As we found that the IMF is correlated with the local stellar mass density, our default models assume 
different
values for these two ingredients. The differences are indicated by model A and B in lines corresponding to the IMF and
local stellar mass density. The remaining model components presented in the third column are the same for the default model 
A or B.
 In subsequent sections, we show how each model ingredient influences the results and by presenting the best fit to data 
for each of them we justify the values chosen to compose our default models. 

 As the investigation of different combinations of IMF and SFR in 
 solar neighbourhood was the central point of our study, it is extensively discussed in Section \ref{dynmass}.
 For other ingredients, we checked two or three different scenarios, while
 the updated ingredients (evolutionary tracks, binarity, atmosphere models, and ISM local density) are systematically considered. 
 The implementation of binarity is undoubtly an important update. The fraction of multiple stars in our Galaxy is 
still unknown, however, most studies suggest that binaries can account for about 50 \% of the total stellar content of the 
Milky Way (\cite{Arenou2011}). Nevertheless, the simulations without binaries are also presented to study the effect of
simulating systems. All analysis were made by comparing star counts and
 $(B-V)_T$ colour distributions between data and simulations.
 For each test, new simulations were performed in one of the 
reference regions presented in Section \ref{tyc}. The simulations within the Galactic plane, at intermediate latitudes,
or at the Galactic pole were done depending on where the expected effect is stronger. Colour 
histograms were plotted always in the same range and with the same number of bins. To evaluate quantitatively
the effect of assigning different values to each ingredient with respect to Tycho-2 data, we have applied 
a  $\chi^{2}$-type statistics test. Two different tests were used to check: to evaluate the adequacy of 
the stellar densities globally and to test the shape of the colour distribution, which is sensitive to the relative star
 densities as a function of mass and age. Considering that $R_i$ stands for the counts in the model histogram and 
$S_i$ for the counts in the Tycho-2 histogram, and $N_{bin}$ is the number of bins. We apply the following:
\begin{enumerate}
 \item Definition of $\chi^{2}$ as used to test the null hypothesis that a given dataset has been drawn from a given 
distribution,
\[   \chi^{2} = \frac{1}{N_{bin}} \sum_{i=1}^k \frac{(R_i - S_i)^2}{S_i}.  \]
With this definition, we are taking the approximation that the Tycho-2 histograms can be used to exactly represent 
actual 
distribution, and, thus, use them as an absolute reference (we assume that the Tycho-2 data set is complete). 
In this case we are checking the absolute number of objects 
per bin with respect to the Tycho-2 data. 

\item Definition of $\chi^{2}$ as used to test the null hypothesis that two datasets have been drawn from the same underlying 
distribution,
\[   \chi^{2} = \frac{1}{N_{bin}} \sum_{i=1}^k \frac{(k_1R_i - k_2S_i)^2}{R_i+S_i} ~~~ \mathrm{where} ~~~ k_1=\sqrt{\frac{\sum_{i=1}^kS_i}{\sum_{i=1}^kR_i}} k_2=\frac{1}{k_1}. \]  
This definition allows to check the similarity of the shape of the two histograms but
 does not take into account the differences in the total number of stars. Thus, this test is of secondary importance but
 helps to analyse the normalized shapes of the resulting $(B-V)_T$ distributions.
\end{enumerate}

Table \ref{Xi2tests} presents the results of both tests for different model ingredients as discussed in the 
following sections. They are divided into three groups: the first group concerns the disc evolution ingredients, 
the second the updated ingredients, and the
third one the additional tests. In the second column, we specify the region of the sky where the simulations
were performed.
 Options (1) and (2) in the third and fourth column stand for two different values of a given model ingredient that are
 compared. In some cases, more than two values were tested; 
however, this table provides a general information about the representative examples and a more detailed comparison is 
explained later on in this section.
To construct Table \ref{Xi2tests} and for the sake of the more detailed analysis presented in the 
following sections, all simulations were performed with the default model B. Only when testing
the extinction models, both default models A and B were checked. 
 For the IMF,  just an example of comparison is presented in the table, as 
it is extensively discussed in Section \ref{sec:looking}. Three different values of the thin disc age were tested, which are presented in two rows. The two values for each ingredient, (1) and (2), are tested
against Tycho-2 data (T).
\begin{table*}
\begin{center}
\begin{tabular}{|c|c|c|c|c|c||c|c|}
\cline{5-8}
\multicolumn{1}{c}{}&\multicolumn{1}{c}{}&\multicolumn{1}{c}{}&\multicolumn{1}{c}{}& \multicolumn{2}{|c||}{The Minimum }&\multicolumn{2}{|c|}{ $\chi^{2}$ test with}\\
\multicolumn{1}{c}{}&\multicolumn{1}{c}{}&\multicolumn{1}{c}{}&\multicolumn{1}{c}{}& \multicolumn{2}{|c||}{$\chi^{2}$ Method}&\multicolumn{2}{|c|}{ scaling factors}\\  
\hline
\multicolumn{1}{|c|}{model}&\multicolumn{1}{|c|}{field}&\multicolumn{1}{|c|}{option (1)}&\multicolumn{1}{|c|}{option (2)}&\multicolumn{1}{|c|}{$\chi^{2}$}&\multicolumn{1}{|c||}{$\chi^{2}$}&\multicolumn{1}{|c|}{$\chi^{2}$}&\multicolumn{1}{|c|}{$\chi^{2}$}\\
\multicolumn{1}{|c|}{ingredients}&\multicolumn{1}{|c|}{}&\multicolumn{1}{|c|}{}&\multicolumn{1}{|c|}{}&\multicolumn{1}{|c|}{(1)-T}&\multicolumn{1}{|c||}{(2)-T}&\multicolumn{1}{|c|}{(1)-T}&\multicolumn{1}{|c|}{(2)-T}\\
\hline
 IMF                            & 1  &   \cite{Kroupa2008} & Kroupa-Haywood v.6    & 140   & 42     & 39     & 14  \\
 SFR                            & 1  & constant            & decreasing (*)        & 171   & 42     & 31     & 14  \\
 thin disc age                  & 1  &    9 Gyr            &   10 Gyr              & 51    & 42     & 16     & 14  \\
                                & 1  &    12 Gyr           &   10 Gyr              & 40    & 42     & 14     & 14  \\
 age-metallicity relation       & 3  & \cite{Twarog1980}   & \cite{Haywood2006}    & 11    & 4      & 6      & 2  \\
 age-velocity relation          & 3  & \cite{Holmberg2009} & \cite{gomez}          & 11    & 4      & 4      & 2  \\
 thick disc                     & 3  & model TkD1          & model TkD2            & 12    & 4      & 3      & 2  \\
\hline  
 binarity                       & 1  & no                  & yes                   & 76    & 42     & 21     & 14 \\
 evolutionary tracks            & 1  & package E1          & package E2            & 50    & 42     & 18     & 14  \\
 atmosphere models              & 2  &  BaSeL 2.2          & BaSeL 3.1             & 23    & 9      & 11     & 3  \\
 ISM                            & 3  & 0.02 $M_\odot pc^{-3}$ & 0.05 $M_\odot pc^{-3}$ & 6 & 4      & 3      & 2   \\              
\hline  
 extinction model:              & & & & & & & \\
 ~~~~~~~~~~~~~~ model A         & GPlane & \cite{drimmel2001}  & \cite{marshall3D2006} & 127 & 418 & 52 & 92  \\
 ~~~~~~~~~~~~~~ model B         & GPlane & \cite{drimmel2001}  & \cite{marshall3D2006} & 153 & 998 & 82    & 143  \\
 radial scale length            & GC &      model SL1      &   model SL2            & 29 & 49  & 17    & 18 \\
                                & GA &      model SL1      &   model SL2            & 49 & 27  & 20     & 15 \\
                                & GR &      model SL1      &   model SL2            & 22 & 22  & 20     & 19 \\
\hline
\hline
\end{tabular}
\end{center}
 \caption{The $\chi^{2}$ tests for eleven model ingredients with respect to the Tycho-2 data. 
  In each case, the simulations were done towards one of the three reference regions: 1-Galactic plane, 
2-intermediate latitudes and 3-Galactic pole or towards a special region. In the case of the extinction models, a bigger 
stripe within the Galaxy plane (GPlane) was chosen for analysis ($|l|<$100 and $|b|<$10). The radial scale length was 
tested towards the 3 cardinal directions: Galaxy centre (GC), anticentre (GA), and rotation (GR).
The symbol (*) is a decreasing SFR from \cite{AumerBinney2009}. }
\label{Xi2tests}
\end{table*}

\subsection{Updating a first set of model ingredients}\label{update1}

\begin{figure} [ht!]
\begin{center}
\includegraphics[width=9cm]{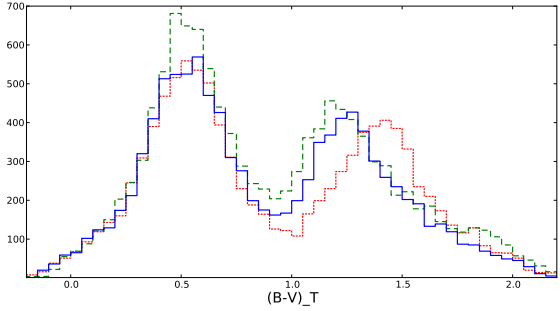}\\
\end{center}
\caption{ Testing atmosphere models. $(B-V)_T$ distributions in the reference 
        region at intermediate latitudes. Tycho-2 data (green dashed line) is compared with
         the simulations obtained with the default model B and BaSeL 3.1 library (blue solid line) 
         and the default model B and BaSeL 2.2 (red dotted line). }
\label{atmos1}
\end{figure}

Here, we discuss how the updates of the atmosphere model, the evolutionary tracks, the binarity treatment, and
the age-metallicity relation have significantly improved the position of the red peak.
In Fig. \ref{atmos1}, we compare simulations using BaSeL 2.2 and BaSeL 3.1 atmosphere model libraries. 
The use of the new BaSeL 3.1 
library moves the red 
peak by more than 0.1 mag towards bluer colours, improving the fit to Tycho-2 distribution, as seen in the figure and 
in the statistics test in the Table \ref{Xi2tests}. We have checked that
the improvement is observed at all 
Galactic latitudes. The MARCS models for giants, when applied, produce a red peak in between the peaks given by the two
 BaSeL libraries, which is not red enough for a good fit to Tycho data. 
From Fig. \ref{atmos1} and the $\chi^{2}$ statistics, it is clear that BaSeL 3.1 makes the BGM fit Tycho-2 data much better.

The update of the stellar evolutionary tracks from those used in \cite{ar2003} to the package E1 and E2 
(Table \ref{tabTracks}) was a necessary improvement  to take into account better physics in stellar models.
When trying both packages, the resulting colour histograms differ slightly in the red peak of the colour 
distribution such that the package E2 leads to 13 \% less stars in the red peak than when package E1 is used. 
 The results of the $\chi^{2}$ tests show a slight improvement when using E2 package, which justifies to use the 
new tracks from \cite{Bertelli2008} and \cite{Bertelli2009} in the default model (package E2).

The age-metallicity relation plays also an important role on the position of the giants peak.
 Table \ref{Xi2tests} shows that the new \cite{Haywood2006} relation clearly reproduces the Tycho-2 data better. 
It performs better in both the blue and the red peak. It is possible that an intermediate 
age-metallicity relation could fit data even better, although the differences in the counts are small in regard to the 
effect of other parameters.
 
\subsection{The dynamical mass and the age-velocity relation}
\label{dynmass}

As shown in Fig. \ref {massModel}, the dynamical mass density and the age-velocity relation (AVR) are the most 
significant input parameters derived from observations that enter into the process of dynamical self-consistency 
(see Block C in Fig. \ref{massModel}). However, the total local dynamical mass is
not directly imposed in the model but it is obtained summing up all mass components. 
If the assumption about the input value of any mass component changes the total dynamical mass also changes. The 
estimation of the local dynamical mass density by \cite{Creze1998} was 
$0.076 \pm 0.015 ~ M_\odot pc^{-3}$. \cite{HolmbergFlynn} derives it to be $0.102 \pm 0.010 ~ M_\odot pc^{-3}$ and 
similarly \cite{Korchagin2003} gives $0.100 ~ M_\odot pc^{-3}$, while  \cite{vanLeeuwen2007} proposes a bit higher 
value $0.122 \pm 0.019 ~ M_\odot pc^{-3}$.
Table \ref{massIngredients} presents the contribution of all mass components to the total dynamical mass density 
obtained here in case of both default models. In italics, we give the components whose density is an input value taken
from external
studies not adjusted in this study.
The thin disc local mass density took two values derived from 
observations 0.039 $M_\odot pc^{-3}$ (LMD1) and  0.033 $M_\odot pc^{-3} $ (LMD2). The best fitted thick disc density 
was set to $2.9 10^{-3}  ~ M_\odot pc^{-3}$ (see Section \ref{add}),  and   
for the stellar halo, we fixed $ 0.92 \times 10^{-5} ~ M_\odot pc^{-3}$. As can be seen in Fig. \ref{massModel}, the local mass 
density of the dark halo is recomputed each time in the iterative process. The values of the local dark matter density
derived in case of model A and B are $ 0.01051 $ and $ 0.01085 $ $M_\odot pc^{-3}$, respectively. They are only 
slightly higher than $0.008 \pm 0.003 ~~M_\odot pc^{-3}$ proposed by  \cite{BovyTrem2012} and 
$0.0075 \pm 0.0021 ~ M_\odot pc^{-3}$, which is recently obtained by \cite{Zhang2013} using SDSS/SEGUE data. Our values are
at less than $2\sigma$ from the values derived by these authors. They also agree with the values 
$0.012 ~ M_\odot pc^{-3}$ derived by \cite{Bienayme2006} from the kinematics and $0.008 ~ M_\odot pc^{-3}$ 
obtained by \cite{deBoer2005} from diffuse Galactic gamma rays for a spherical dark halo. This parameter is still 
much uncertain as discussed in \cite{Burch}. These authors perform a self-consistent calculation of the spatial 
distribution of dark matter and try to contrain it by comparing it with kinematic observables. They find the local dark
matter density to be greater than the recent estimates: $0.015$ and $0.019 ~~M_\odot pc^{-3}$.
Depending on which local stellar volume mass density was assumed for the thin disc, the total dynamical mass 
density slightly changes. However, by taking into account the error bars, both values are in a good 
agreement with the results of \cite{HolmbergFlynn} and \cite{Korchagin2003}. Undoubtedly, the derived values are highly
dependent on the density assumed for the ISM (see Section \ref{sec:LMD}).
\begin{table}
\begin{center}
\begin{tabular}{|c|c|c|}
\hline \hline \hline
\multicolumn{1}{|c|}{mass ingredients}&\multicolumn{1}{|c|}{model A}&\multicolumn{1}{|c|}{modelB}\\
\hline
{thin disc 1}       & 0.00196 & 0.00188\\
 ~~~~~~~~~~~~~~~{2} & 0.00545 & 0.00504\\
 ~~~~~~~~~~~~~~~{3} & 0.00464 & 0.00411\\
 ~~~~~~~~~~~~~~~{4} & 0.00333 & 0.00284\\
 ~~~~~~~~~~~~~~~{5} & 0.00582 & 0.00488\\
 ~~~~~~~~~~~~~~~{6} & 0.00609 & 0.00502\\
 ~~~~~~~~~~~~~~~{7} & 0.01169 & 0.00932\\
  \textit{total thin disc}  & 0.039 & 0.033\\
\textit{white dwarfs} & 0.00714 & 0.00714  \\
 \textit{thick disc} & 0.00291 &  0.00291\\
\textit{stellar halo} & 0.92 $\times 10^{-5}$ & 0.92 $\times 10^{-5}$ \\
\textit{ISM} & 0.05 & 0.05 \\
{dark halo} & 0.01051 & 0.01085  \\
\hline
   Total & 0.10954  & 0.10399 \\
\hline
\end{tabular}
\end{center}
\caption{Contribution of all mass components (in $M_\odot pc^{-3}$) to the total dynamical mass in the default 
models A and B. In italics, we specify the fixed parameters:  thin disc total stellar mass density and the density of
white dwarfs, thick disc, stellar halo and ISM. }
\label{massIngredients}
\end{table}

\begin{figure} [h!]
\begin{center}
\includegraphics[width=9cm]{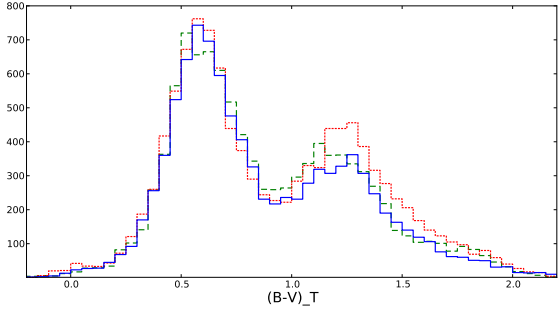}\\
\end{center}
\caption{ $(B-V)_T$ distributions towards the Galactic north pole showing  the effect of changing the age-velocity 
dispersion relation (AVR) against the default model B. 
 Tycho-2 data (green dashed line) and the simulations based on the default model B (blue solid line) are 
shown. The red dotted line is based on simulations using an AVR of \cite{Holmberg2009} 
rather than of \cite{gomez} but leaving all other parameters unchanged. }
\label{ll4}
\end{figure}

The age-velocity dispersion relation (AVR) investigation is presented in Fig. \ref{ll4} and Table \ref{Xi2tests}. It shows
 that the \cite{gomez} AVR leads to better results at the 
Galactic pole than the \cite{Holmberg2009} AVR. The latter produces too many giants and bright stars within the Galactic plane. 

\subsection{Ingredients for star generation in mass and time }
\label{masstime}

As expected, the IMF, SFR, and the local stellar volume mass density are very correlated; this is why we studied
these three ingredients collectively. In this section, we also discuss the effect of binarity on the simulations.
Different combinations were tested using two approaches. The first one was 
to compare the synthetic LF produced
by the given combination with the observed LF. The second step was to check the fitting of the obtained simulations
to Tycho-2 data.

\subsubsection{Comparison with the LF}

When fitting the local LF, we have considered the total LF from 
\cite{Jahreiss1997} data. The procedure of comparing the synthetic and observed LFs 
was repeated for all tested combinations of eleven IMFs, four SFRs (Sect. \ref{sec:imfsfrRes}) and two values of the 
local stellar volume mass 
density (Sect. \ref{sec:LMD}). For some of the combinations, we also have performed a secondary test that excludes
the giants 
from the 
simulations and compares them to the \cite{Kroupa2001} and \cite{Reid2002} luminosity function of main sequence stars.

Using both values of the local stellar mass density, the LMD1 and LMD2, the constant SFR was tested along with all eleven IMFs proposed in Table 
\ref{tabIMF}.  
Most IMFs combined with a constant SFR produce many more bright objects than the observed LF, except for two of them, 
the Haywood-Robin and \cite{Kroupa2008} IMF, for which the solution agrees with the 
observational LF from \cite{Jahreiss1997}. As shown in Section \ref{sec:FitTycho}, Tycho-2 data turns out
to be an even stronger constraint when evaluating constant SFR.

Secondly, all IMFs were combined with a decreasing SFR and both values of the local stellar volume mass density. 
In most cases when using the LMD2, the simulated LFs have a better fit to data than when we assume the LMD1. 
Some of the combinations led to a good fit in one range of magnitudes and a bit worse in the other. This is 
the case of \cite{Kroupa2008} IMF. This function fits well the observed LF in the range of bright magnitudes. 
However, in the region of $4 < M_V < 8$ mag, the default models A and B give a much better fit than the combination with
 \cite{Kroupa2008} IMF. 
The \cite{Vallenari2006} IMF combined with a decreasing SFR and two values of the local stellar volume
mass density always produces a LF above the observational values.

\begin{figure*} [ht!]
\begin{center}
\includegraphics[width=12cm, angle=-90]{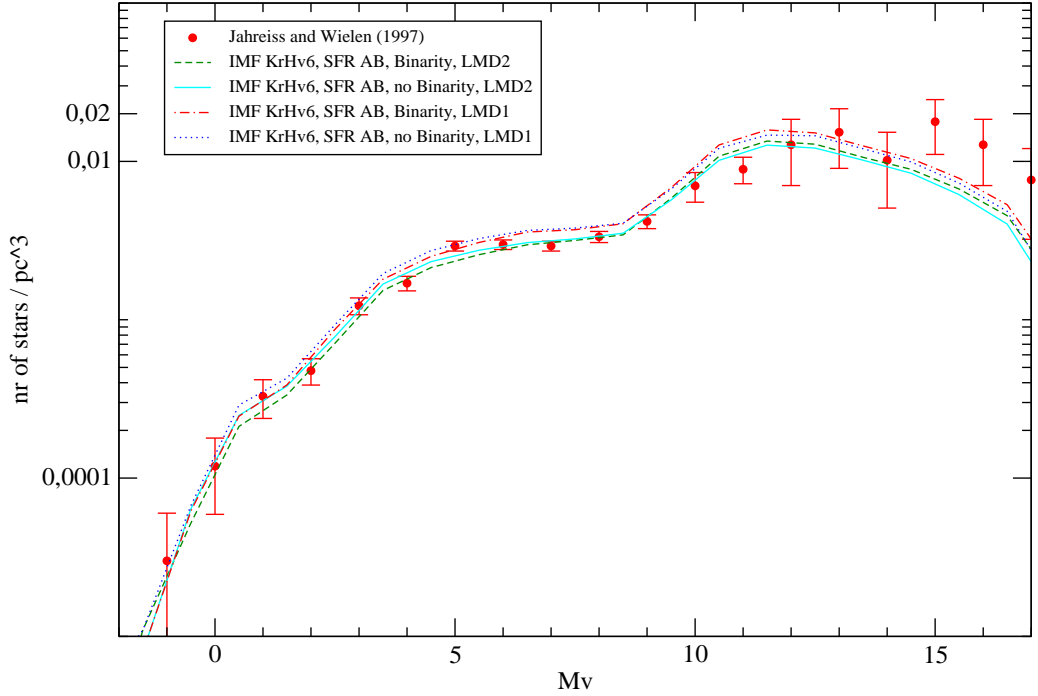}
\end{center}
\caption{Comparing the observed LF with the synthetic LF that is obtained with the default model B and two values of local 
         stellar mass density, LMD1 and LMD2 (in both cases simulations with and without binaries). Total LF from 
         \cite{Jahreiss1997} (red dots), main sequence LF from \cite{Kroupa2001} (black triangles), and 
         \cite{Reid2002} (black dots) are shown.}
\label{rhoObs3}
\end{figure*}

In Fig. \ref{rhoObs3}, we present the comparison of the 
observed LF with the synthetic LF, as obtained with the default 
model B (Table \ref{tableParam}) and two values of the local stellar mass density, LMD1 and LMD2 (in both cases simulations with and without
binaries). As expected, changing the value of the local mass density from  LMD1 to LMD2 shifts the whole LF downward. 
Binarity produces a second order effect, which is discussed in the comparison with Tycho data (Sect. \ref{sec:FitTycho}).

\begin{figure} [ht!]
\begin{center}
\includegraphics[angle=-90,width=8.5cm]{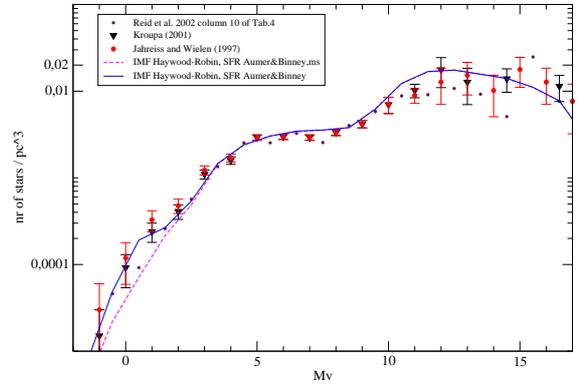}
\end{center}
\caption{Comparison of the observed LF with simulated one from model A: total LF (blue solid line) and main sequence LF 
(magenta dashed line).  Symbols for observed data are same as in fig. \ref{rhoObs3}. }
\label{newModel2}
\end{figure}

In Fig. \ref{rhoObs3}, it is shown that the default model B
with the Kroupa-Haywood v6 IMF and a decreasing SFR (\cite{AumerBinney2009}) reproduces better the local LF in the 
range of absolute magnitude [4;12] when using the LMD2 than when the LMD1 is applied. At the bright tail of the 
LF, both solutions are within the big error bars. 
Performing the same comparisons with the default model A, LMD1 provides a LF, which is closer to the observations. 
TIn the following, we consider two default models, the model A (Haywood-Robin IMF) with LMD1 and model B (Kroupa-Haywood v6 IMF) with 
LMD2. Figure \ref{newModel2} shows the comparison of the total and main-sequence LFs produced by the default
model A with the observations.

Taking into account
big error bars at the bright and faint ends of the observed LF, we can conclude that most of the 
model LFs do not differ much from the observations. However, in the following section we show 
that this conclusion arises from the uncertainties on the local luminosity function mainly due to Poisson statistics 
for bright stars. In next section we show that Tycho-2 data gives stronger constraints and are able to distinguish between several hypothesis on the
IMF. As an example, the default model with IMF changed to the one of \cite{Scalo1986} and the LMD2 local density fits well 
 the \cite{Jahreiss1997} LF for $ M_V < 8$ mag, but it does not fit well the Tycho-2 $(B-V)_T$ distribution.
 Most of the Tycho-2 sample covers the range of absolute 
 magnitude $M_V=[-1:5]$. It includes the region where the LF has large error bars ($M_V<3$). 

\subsubsection{Comparison with Tycho-2 data}
\label{sec:FitTycho}

\begin{figure} [ht!]
\begin{center}
\includegraphics[width=9cm]{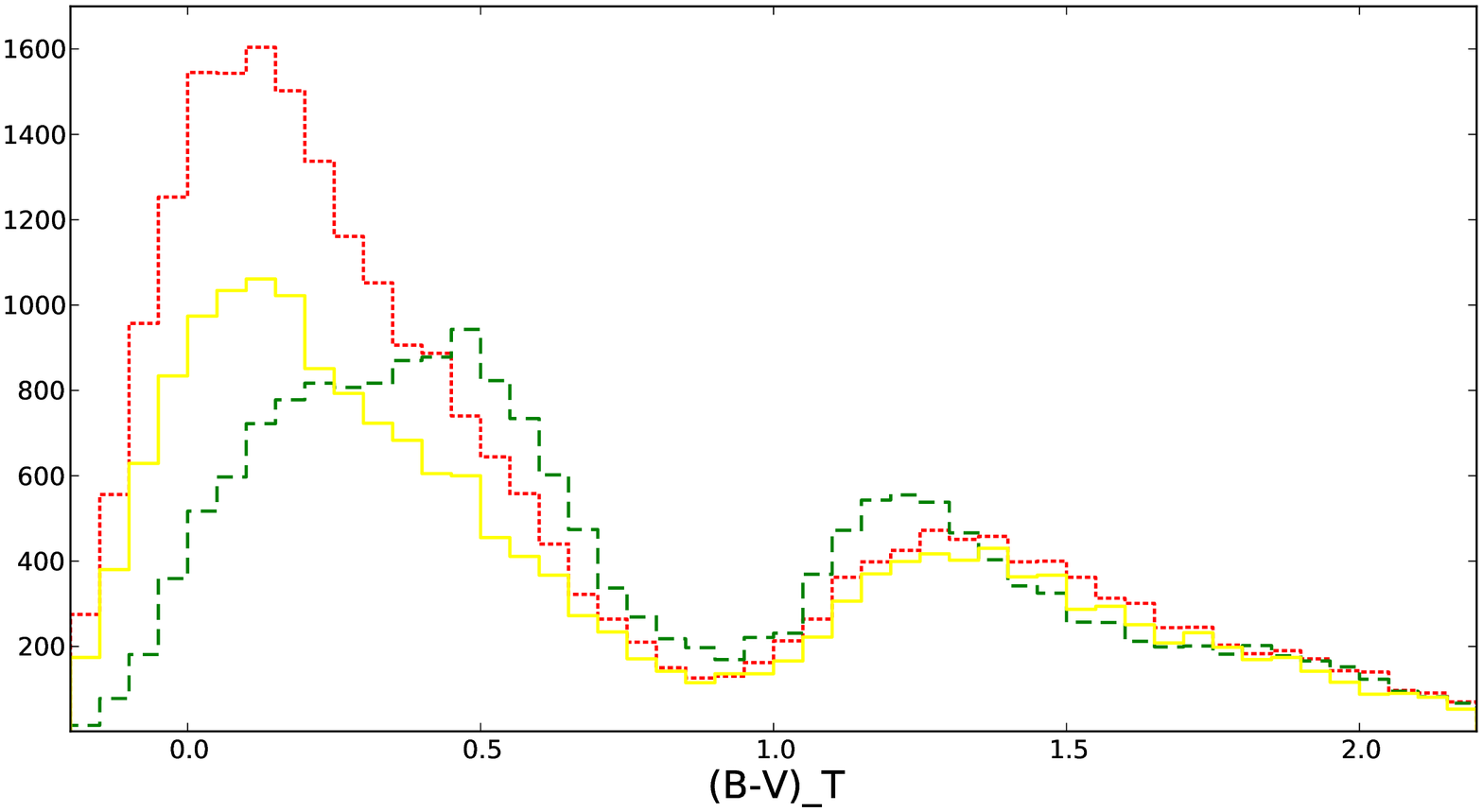}\\
\includegraphics[width=9cm]{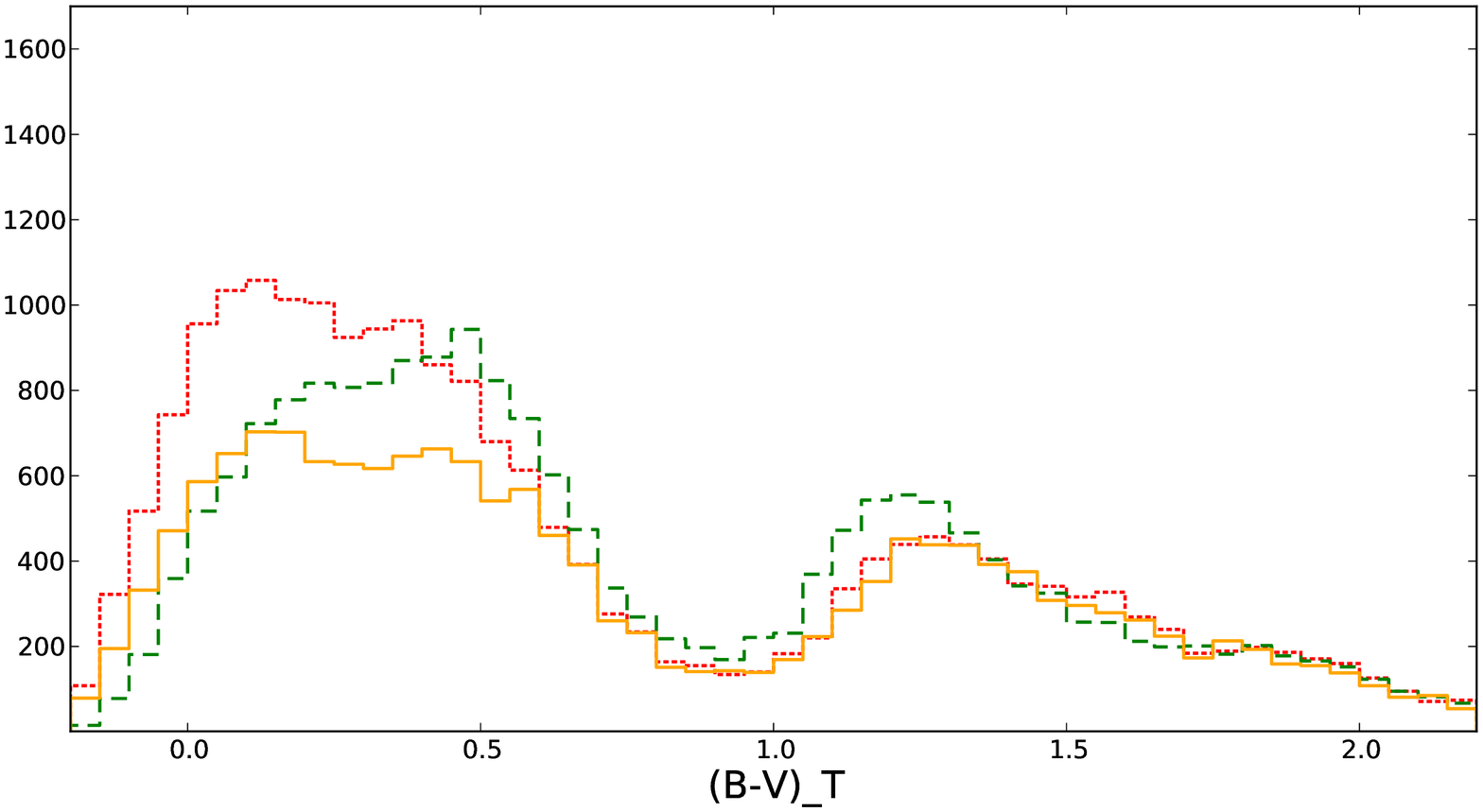}\\
\includegraphics[width=9cm]{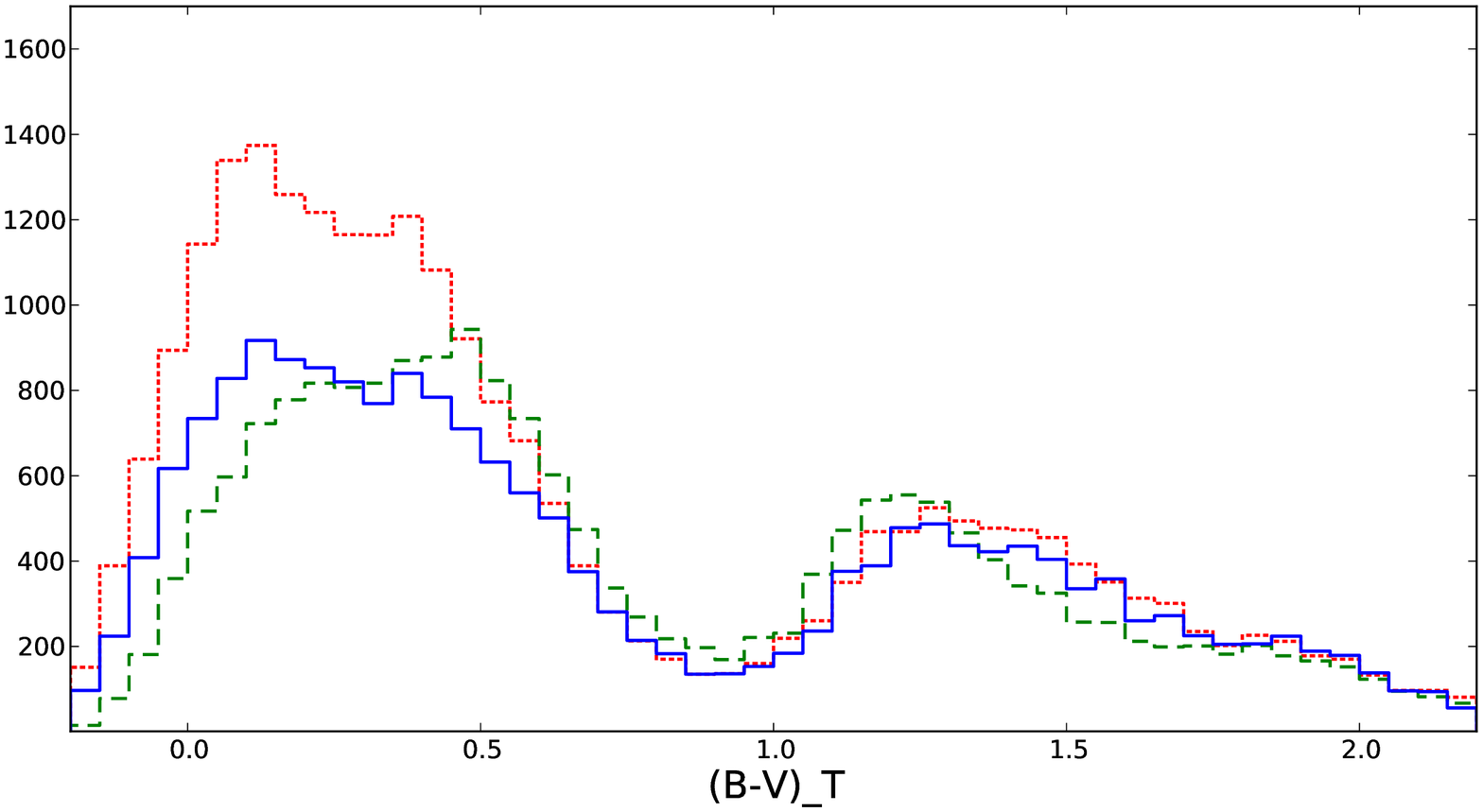}\\
\end{center}
\caption{Testing a constant and decreasing SFR with different IMFs. All three graphics show
 the $(B-V)_T$ distributions of the reference region within the Galactic plane. Tycho data in all panels is plotted using
 green dashed line.
The top panel corresponds to \cite{Kroupa2008} IMF, the middle to Haywood-Robin IMF and the bottom 
to Kroupa-Haywood v6 IMF (see Table \ref{tabIMF}). In red dotted line the simulations, which are obtained with constant SFR, 
and in yellow, orange, and blue solid lines, the simulations obtained with the decreasing SFR of \cite{AumerBinney2009} and 
each IMF respectively. }
\label{gggc}
\end{figure}

The $(B-V)_T$ colour distributions of Tycho-2 data allow us to distinguish the blue main sequence stars from
the giants, while they are both at about the same absolute magnitude in the LF. Figure \ref{gggc} presents the 
colour distributions when three IMFs were combined with a constant and a decreasing SFR. When \cite{Kroupa2008} and 
Kroupa-Haywood v6 IMFs are combined with a constant SFR, many more blue 
(on average young) objects are produced than when a decreasing SFR is applied. In the case of the Kroupa-Haywood v6 
IMF and both SFRs, we have performed a $\chi^{2}$ statistics presented in Table \ref{Xi2tests}. There is a clear improvement
 of the fit when assuming a decreasing SFR. Only when the Haywood-Robin 
IMF is combined with the constant SFR, the excess of the blue stars is small. For all 
other IMFs that we have tested, the excess of blue stars is very significant and, in some cases, even bigger  
than presented in Fig. \ref{gggc} (for example \cite{Scalo1986} and \cite{Vallenari2006} IMF). We conclude that we are 
getting unrealistic star counts on the blue part of the $(B-V)_T$ distribution within the Galactic plane when imposing
a constant SFR  along with eleven IMFs that we tested. Only in the case of  
Haywood-Robin IMF, this excess of stars is less significant. Tycho data strongly favour a decreasing SFR.
This is most likely one of the main reasons of the disagreement between the old model and Tycho data.  

Then, we have combined all tested IMFs with the decreasing SFR from \cite{AumerBinney2009}. The IMF from \cite{JJ2010} was additionally combined with their 
best fit SFR (see model A from their paper). We refer this combination of SFR and IMF as the \cite{JJ2010}
best fit. The details of the analysis made for all IMFs are presented in \cite{praca}. Many of the combinations
led to a significant excess of the blue and bright stars within the Galactic plane  
(in particular \cite{Scalo1986}, \cite{Kroupa2008}, \cite{Vallenari2006} and KrH v8  IMFs). Other IMF produced far too few
 objects (KrH v1, KrH v4 and \cite{misha1997-1}).

\begin{figure}[ht!]
\begin{center}
\includegraphics[width=9cm]{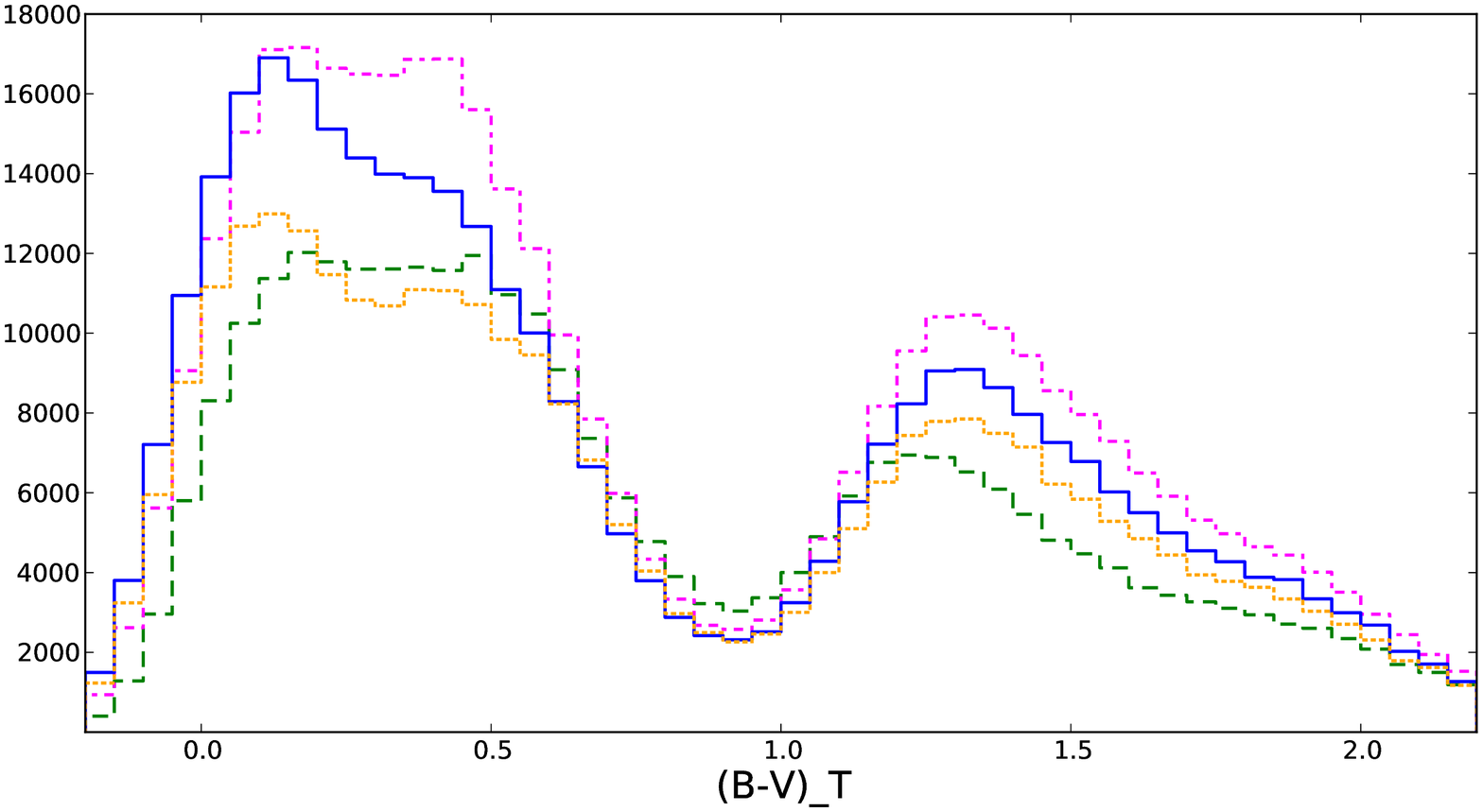}  \\
\includegraphics[width=9cm]{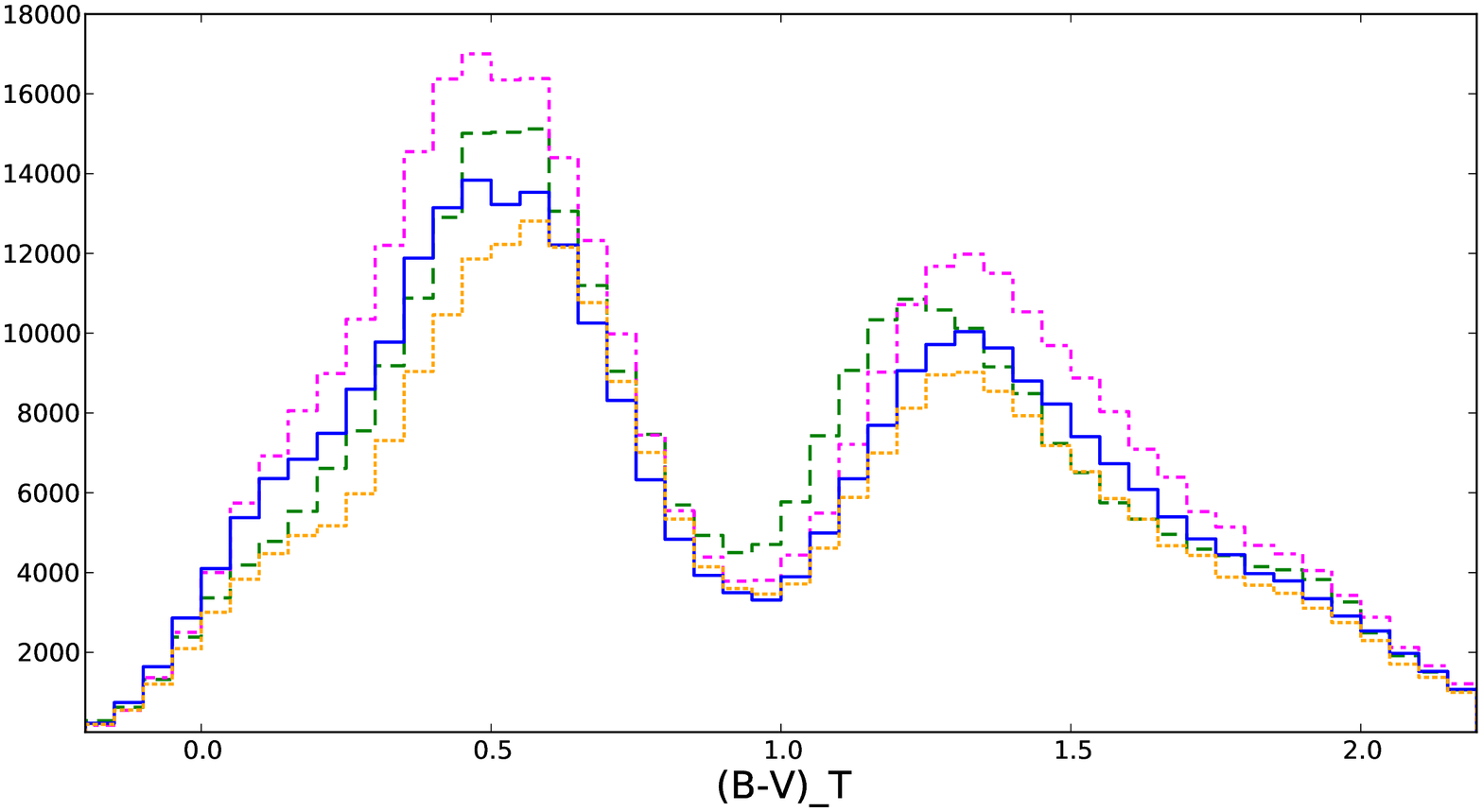} \\
\includegraphics[width=9cm]{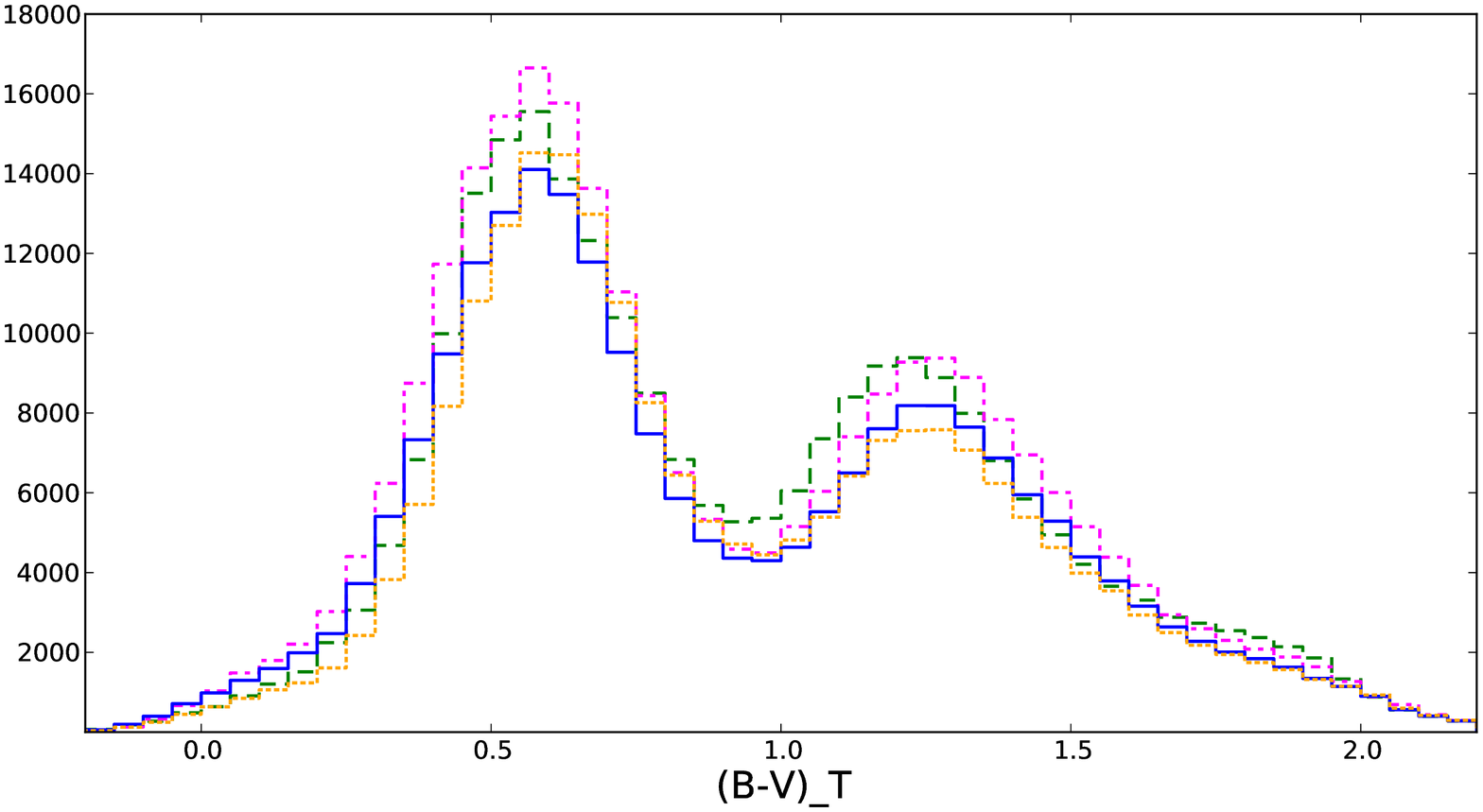}\\
\end{center}
\caption{Testing different IMFs. The $(B-V)_T$ distributions correspond to the (top) entire Galactic plane $|b|<$10, 
(middle)
 entire stripe of the sky at 10$<|b|\leq$30, and (bottom) the stripe of the sky at 30$<|b|\leq$90. 
 In orange dotted line the simulations, which are obtained with the default model A (Haywood-Robin IMF); in blue solid line
 we have the default model B (Kroupa-Haywood v6 IMF), and the default model combined with the best fit SFR and IMF 
from \cite{JJ2010} is in magenta dashed-dotted line. The Tycho-2 data is in green dashed line.}
\label{whs2}
\end{figure}

The best fit was obtained 
when Haywood-Robin and Kroupa-Haywood v6 IMF were combined with a decreasing SFR of \cite{AumerBinney2009}.
These two combinations were used in default models A and B respectively (see Table \ref{tableParam}).
 The best fit SFR and IMF from \cite{JJ2010}, as combined with the default model (for the rest of the 
ingredients) (hereafter the JJ model) also provides a good solution. In Fig. \ref{whs2}, we show the whole sky
$(B-V)_T$ distributions obtained with the two default models
 and the JJ model. These distributions were divided
 into three latitude ranges corresponding to Galactic plane, intermediate latitudes and the Galactic poles.
 At the intermediate latitudes, the best fit is provided by the model B, and at the Galactic poles, both the model B 
 and the JJ model are the best.
 It can be seen that Haywood-Robin IMF (model A) fits the star counts of Tycho-2 within the Galactic plane the best.
 However, the fit provided by model B is one of the best when compared to all other tested IMFs. The 
   JJ model produces more stars than the model B.
  When looking at the Galactic plane, we can notice that there are still some discrepancies present. 
  The Galactic plane is where many different factors interplay and finding a good fit is a very complex task 
  (see Sect. 5.6).
  \begin{figure} [ht!]
\begin{center}
\includegraphics[width=9cm]{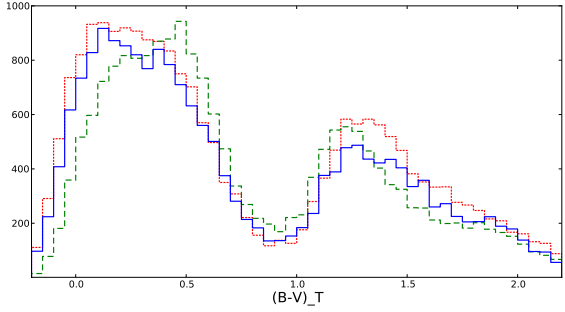}\\
\end{center}
\caption{ Testing binarity. $(B-V)_T$ distribution in our 
reference region within the Galactic plane. Tycho-2 data (green dashed line), the simulations obtained with the default 
model B and binarity (blue solid line) and simulations with the default model B without binarity (red dotted line) are shown.  }
\label{pp1}
\end{figure}

In Fig. \ref{pp1}, we present the simulations with and without binaries. As expected, 
when adding binarity, we increase the number of objects at the low-mass tail of the LF and decrement the high masses, as 
 also seen on the LF in Fig. \ref{rhoObs3}.
One notices the decrease of star counts in the simulations with binaries in Fig. \ref{pp1}. This is because the 
stars up to magnitude $V_T \leq 11$ correspond to the bright end of the LF. 
As can be seen in this figure, the simulations with binaries give a better fit to Tycho-2 distribution, although we have 
to take into account that we are testing this parameter here independently from the others. We have also checked 
the number of produced binaries in the whole sky simulations. As expected, the percentage of created
binaries is the same for both default models. In the whole sky sample, 45 \% of stars are unresolved binaries, 12 \% are 
resolved systems, 
and 35 \% are single stars (the rest are thick disc stars).

\subsection{Additional tests}
\label{add}

In this section, we discuss more ingredients that are critical within the Galactic plane. Work is in progress to constrain
 them using additional data sets.

\begin{figure} [ht!]
\begin{center}
\includegraphics[width=9cm]{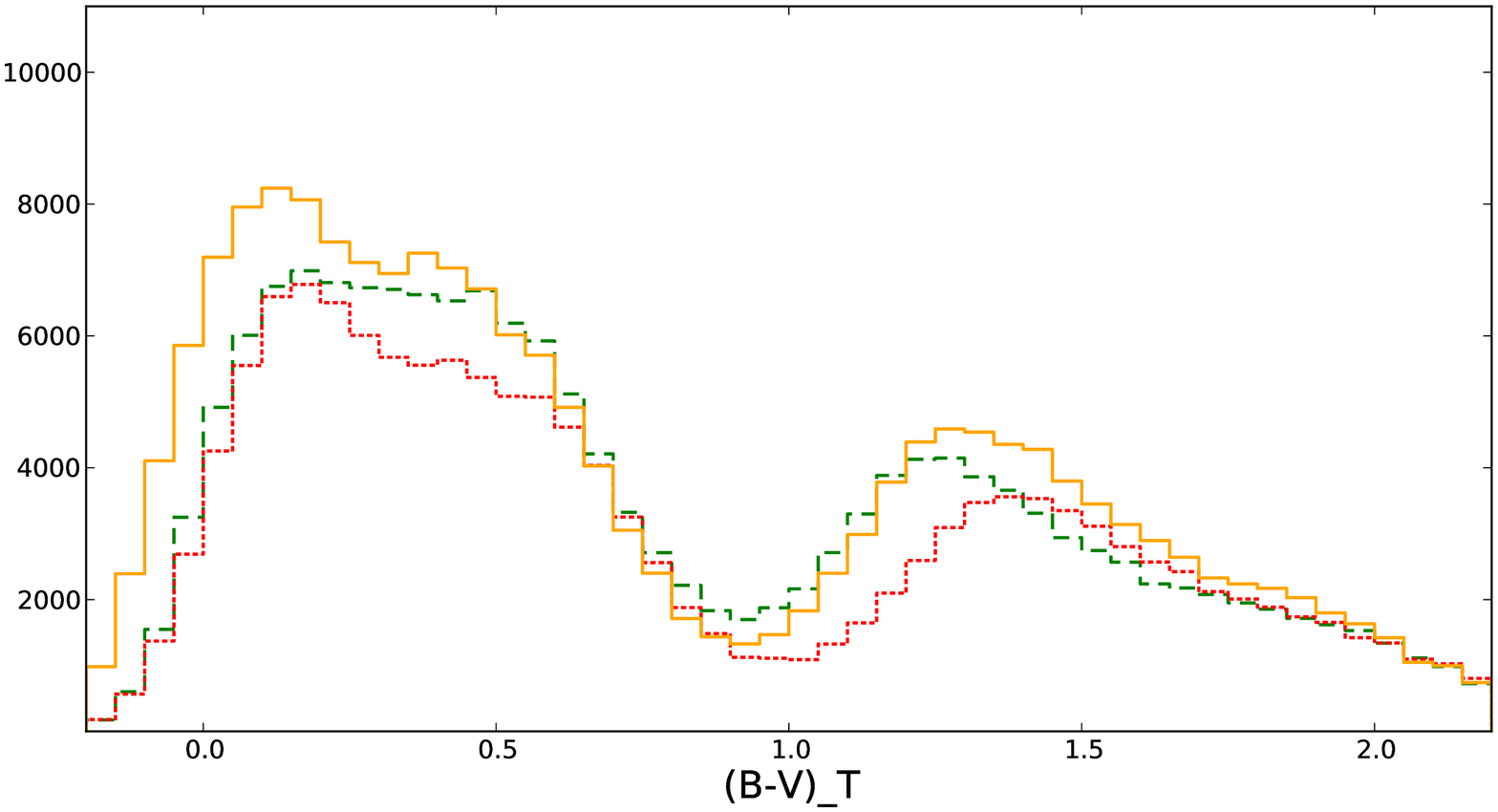}\\ 
\includegraphics[width=9cm]{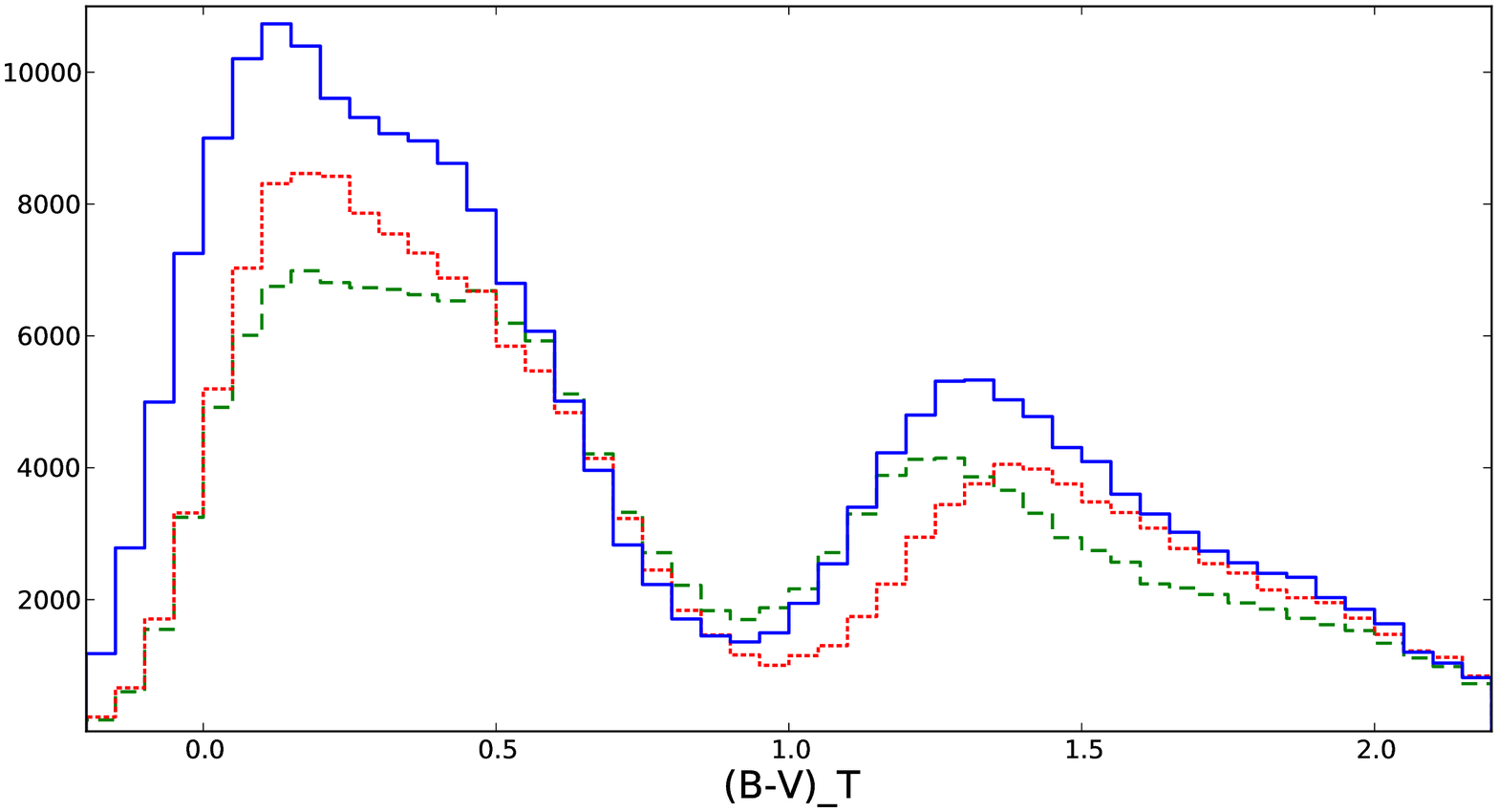}\\ 
\end{center}
\caption{ Testing two different extinction models. The presented $(B-V)_T$ distributions correspond to a part of
the Galactic plane $|l|<$100 and $|b|<$10. At the top of default model A and at the bottom of default model B, two 
different extinction models are also used: dotted red line is \cite{drimmel2001} (in both cases), orange solid line is
 model A with \cite{marshall3D2006}, blue solid line is model B with \cite{marshall3D2006} and dashed green is Tycho data.}
\label{ext1}
\end{figure}

 Two extinction models (Sect. \ref{sec:extinction}) were compared in the stripe at $|l|<$100 and $|b|<$10.
 As can be seen in Fig. \ref{ext1}, the difference in star counts is significant. 
It is possible that the \cite{marshall3D2006} extinction model underestimates the absorption at the very 
short distances, what could be caused by the lack of data in the local sphere (they used K giants from 2MASS and 
these stars are very few at short distances).
This could explain the excess of blue stars. On the contrary, the \cite{drimmel2001} extinction model produces 
 a deficiency of giant stars and a shift of their peak to redder colours. This could be explained by an overestimate
 of the extinction at large distances. From the $\chi^{2}$ test (Table \ref{Xi2tests}), it seems that the 
 general fit obtained with \cite{marshall3D2006} extinction model is worse, although it performs 
 better in the red peak. Changing the extinction model in only the Galactic plane stripe $|l|<$100 and $|b|<$10 leads to a
 change in the all sky star counts, which reaches  $\sim$ 10 \%  for the bluest and bright objects ($(B-V)_T < $ 0.4).
As expected, the extinction model is a crucial input of BGM. However,  
several parameters 
influence the star counts within the plane. Complementary analysis would be needed to derive any sensible conclusions.

 We now explore the effect of the age adopted for the thin disc. It appears that the effect is sensitive to different values
when we look at different latitudes. At the Galactic pole, the samples simulated with 9, 10, and 12 Gyr behave similarly in
 the blue peak, while the solution obtained with 12 Gyr produces more old stars and 
fits better Tycho-2 data in the red peak. This is due to the fact that older stars have a larger scale height following the 
increase of the velocity dispersion. At intermediate latitudes, the same effect is visible but then the number
of giants is larger than in Tycho sample. At these latitudes the best solution is provided by 
simulations with a 9 Gyr  or 10 Gyr old thin disc. In the Galactic plane, the difference between the three ages are 
very small, as seen in the $\chi^{2}$ tests 
(Table \ref{Xi2tests}). The differences are mostly observed in the blue peak, while the giant peak is nearly insensitive to
 this parameter.  If one extends the 
age of the thin disc to 12 Gyr, stars that are a bit less blue and young are produced within the plane, which makes the fit to 
Tycho-2 data slightly better. However, the Galactic plane is a problematic region, where extinction, spiral arms and  
many other effects can play and compensate the effect of the age. Hence, we keep the conservative value of 10 Gyr, which 
provides the best fit to intermediate and high latitudes. 

The Tycho-2 data do not reach large distances (from the solar neighbourhood) giving relatively poor constraints on the
 radial scale length, as compared to more remote star counts. 
The $\chi^{2}$ tests in Table \ref{Xi2tests} show that the change, as expected, 
is negligible towards the rotation but is significant 
towards the Galactic centre (main sequence) and anticentre (giants). As the new scale length improves in one case and
degrades in the other, it might be due to another effect, which plays at low latitudes. We thus keep the default value. 

Finally, two different sets of thick disc density parameters (section \ref{sec:thick}) have been tested in
simulations and compared to Tycho data.
As we are limited to $V_T$=11, the thick disc provides a small contribution to the whole sky statistics: 
the relative number of thick disc stars in our sample is 7.3 \% for the \cite{Reyle2001} best fit model and only
2 \% in the other case.   The model that assumes a larger local density for the thick disc provides a better fit to 
Tycho-2 data, as shown by the $\chi^{2}$ tests, in Table \ref{Xi2tests}. 
It fits better the red giants peak and significantly contributes at intermediate 
colours in between the peaks. Deriving a firm conclusion about the thick disc parameters requires a specific analysis of large 
scale surveys.

\subsection{Looking for the best fit with Tycho-2 data}\label{sec:looking}

\begin{figure}
\begin{center}
\includegraphics[width=9cm]{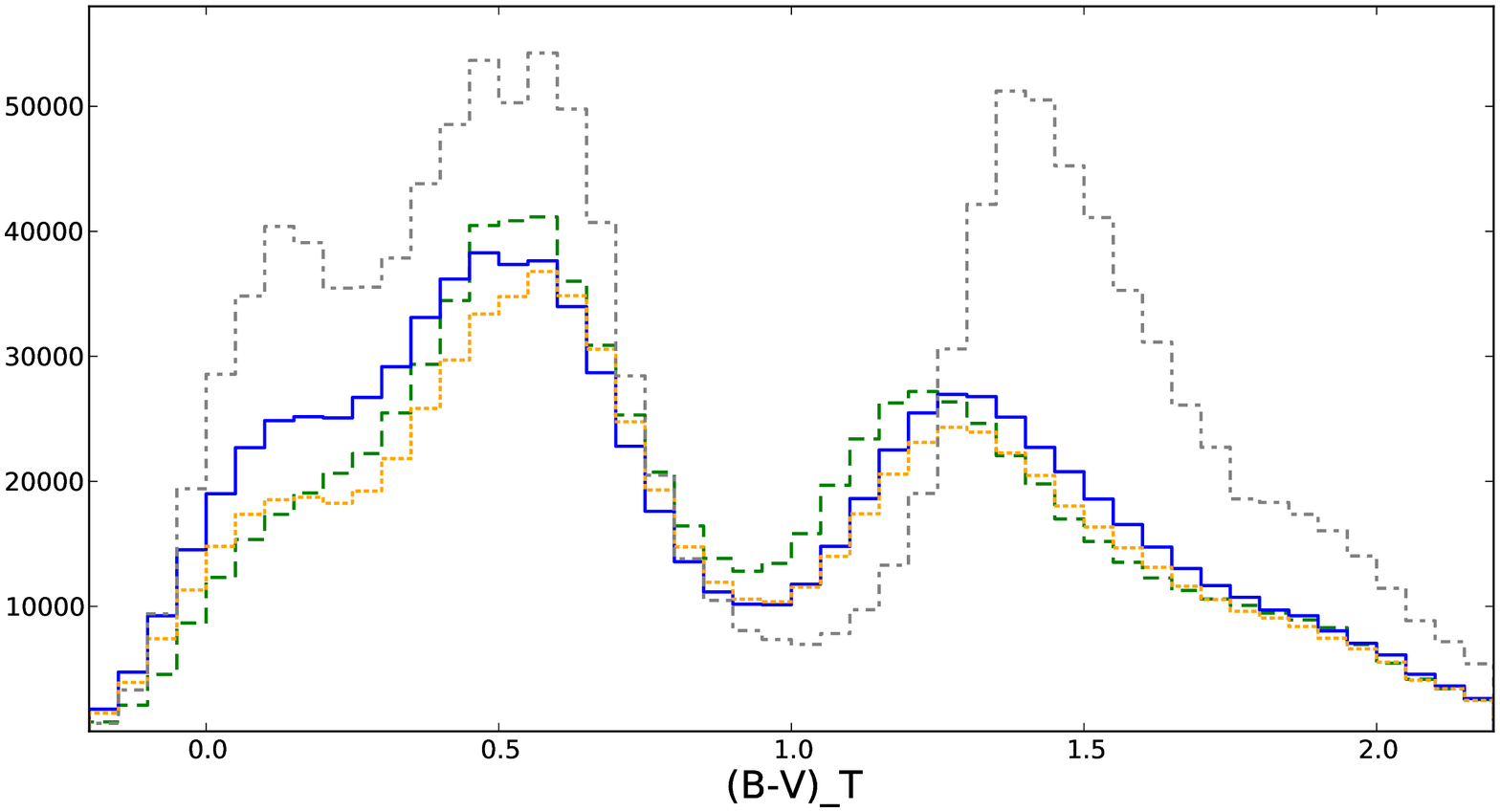}\\ 
\end{center}
\caption{Whole sky $(B-V)_T$ distributions of the 
old model (grey dashed-dotted line), Tycho-2 data  (green dashed line), and two new default models: model A in orange dotted
line (Haywood-Robin IMF) and model B in blue solid line (Kroupa-Haywood v6 IMF).}
\label{newModel}
\end{figure}

In Fig. \ref{newModel}, we present the whole-sky $(B-V)_T$ distributions of the two new default models, the old model
 and Tycho-2 data. Notice the enormous 
changes in the $(B-V)_T$ histogram of the new model with respect to the old one: we have significantly
 decreased the number of 
objects in both peaks making it in much better agreement with the Tycho-2 sample. The red peak of giant stars has been
 shifted by more than 0.1 
mag towards the blue colours, which fits significantly better the giant peak of Tycho-2 data. The excess of the blue stars around
 $(B-V)_T \sim$ 0.1 and 0.5 mag has been significantly reduced. Table \ref{tableNRObjects} summarizes the observed 
star counts and the behavior of the models in different latitude ranges.
 
From Fig. \ref{whs2}, we also derive the conclusions that the default model A (Haywood-Robin IMF) 
reproduces the $(B-V)_T$ distribution within the Galactic plane very well, while it produces, respectively,
 15 and 12 \%  less stars than Tycho-2 data at intermediate and higher latitudes.
The default model B (Kroupa-Haywood v6 IMF) differs from the data more significantly at the Galactic plane but
it performs very well at intermediate and high latitudes.
 
\begin{figure}
\begin{center}
\includegraphics[width=9cm]{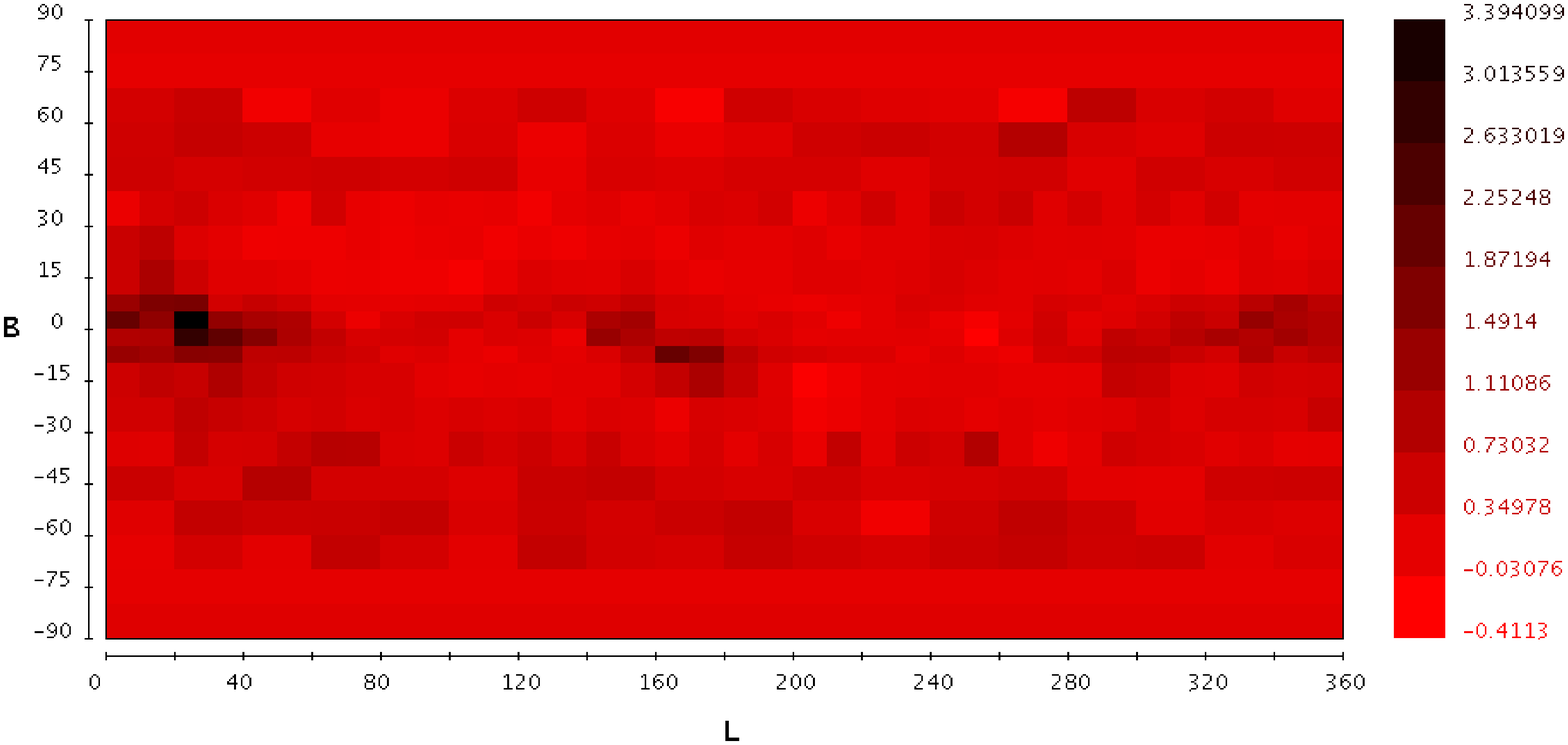}\\
\includegraphics[width=9cm]{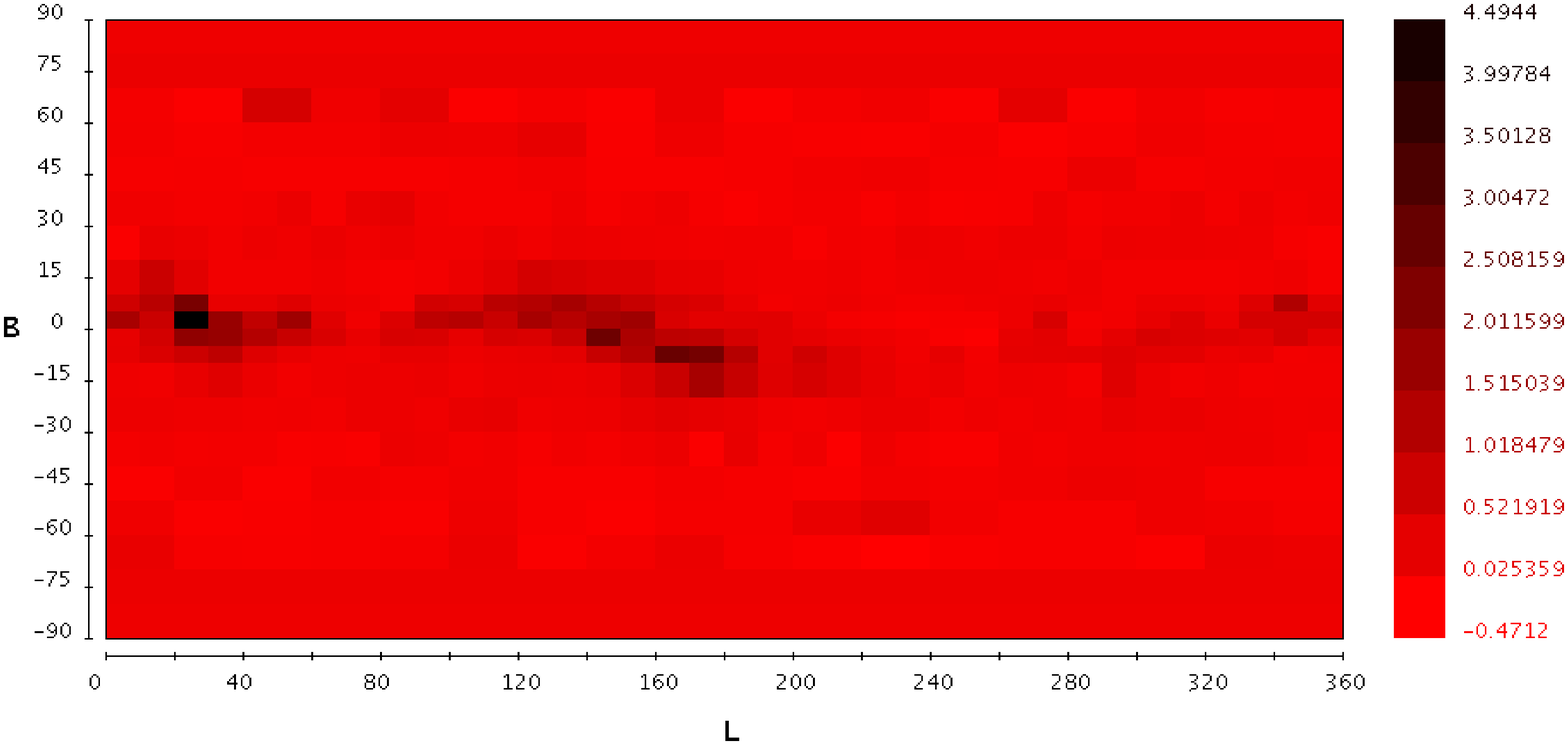}\\
\end{center}
\caption{ Sky maps of
the relative number of objects (model B -Tycho)/Tycho between model B and Tycho data for two colour groups as defined by a 
clustering technique. Top blue group, bottom: red group. }
\label{newModel-3}
\end{figure}

 As a complementary test, a clustering technique was performed on the $(B-V)_T$ distributions to separate the
sample into two groups, the blue and the red. Figure \ref{newModel-3} shows sky maps of the relative differences $(N_{Model}-N_{Tycho})/N_{Tycho}$ in the number of 
objects (whole sky, default model B) for the blue group (top) and the red group (bottom). As expected, the Galactic plane 
is the most problematic region due to high
interstellar extinction and its somewhat clumpy distribution. The largest discrepancies in the relative
number of objects are correlated with the discrepancies in the mean colours. The 
regions within the plane where there are fewer Tycho-2 star counts than in the model are also clearly seen in Fig.1 from
 \cite{tycho2}. The two most problematic regions can be associated 
with dust clouds, the Taurus dark cloud at l=174${^\circ}$ and 175 pc and the large complex Aquila rift between 
l=10${^\circ}$ and l=55${^\circ}$ that runs from 50 pc to 375 pc \citep{lucke}. The 
extinction models used here are not realistic in those close dark clouds. Some other structures not considered
in the comparisons can interfere as well, such as a spiral structure, star-forming regions, Gould belt, etc. In spite of these localized 
discrepancies, we have significantly improved  (by at least a factor of two) the accuracy of star count predictions obtained with the 
new model when compared with old model \citep{praca}. 

In Table \ref{tableChi}, $\chi^{2}$ tests are applied to the whole sky distributions.
The most striking result from this table is that
the new default models A and B and the JJ model provide a huge 
improvement with respect to the old model. The most significant improvement is observed for $|b|>10$, while
the Galactic
plane is the most problematic region, where the modelling of the extinction causes a large part of the discrepancies.
According to the second $\chi^{2}$ test with scaling factors, the model A is best at all latitudes,
meaning that the shape of the distributions produced by this model is the closest to data.
 In absolute terms (the minimum $\chi^{2}$ test), model B fits better the observed data
for all the sky outside the plane ($|b|<10$). The results obtained with JJ model at high latitudes are also very good  in terms of absolute star counts
and the distribution's shape. As for model B, it becomes less realistic within the Galactic plane.

  Despite of the huge improvement with respect to the old model it is clear that, the simulations 
differ from the data by much more than expected from purely Poissonian statistics, which accounts for all remaining 
uncertainties in the model inputs, including stellar physics, interstellar extinction, inhomogeneties and non-axisymetries 
in the Galaxy components. We look forward to continue to improve the model by further investigations.
 
\begin{table}
\begin{center}
\begin{tabular}{|c|c|c|c|c|}
\hline \hline \hline
\multicolumn{1}{|c|}{}&\multicolumn{1}{|c|}{Tycho}&\multicolumn{1}{|c|}{}&\multicolumn{1}{|c|}{}&\multicolumn{1}{|c|}{ }\\
\multicolumn{1}{|c|}{Region}&\multicolumn{1}{|c|}{ counts}&\multicolumn{1}{|c|}{Model A}&\multicolumn{1}{|c|}{Model B}&\multicolumn{1}{|c|}{Old model}\\
\hline
 &  &   &  &  \\
 all sky & 860743  &  8 \% less & 4 \% more & 50 \% more \\
         &         &  &  &  \\
\hline
  &  &   &  &  \\
 $|b|<10$ & 288661 & 5 \% more  & 23 \% more & 61 \% more \\
                                 &        &            &  &   \\
\hline
&  &  &  &  \\
  10$<|b|<$30 & 322857 & 15 \% less  & 4 \% less  & 15 \% more  \\
              &       &  &  &   \\
\hline
&  &  &  &  \\
 $30<|b|<90$ & 249225 & 12 \% less  & 8 \% less & 37 \% more \\
             &       &  &  &  \\
\hline
\end{tabular}
\end{center}
\caption{ Star count comparisons between model A, model B, and the old model with respect to Tycho-2 data for $V_T\leq$ 11 in three latitude ranges.}
\label{tableNRObjects}
\end{table}
\begin{table*}
\begin{center}
\begin{tabular}{|c|c|c|c|c||c|c|c|c|}
\cline{2-9}
\multicolumn{1}{c}{}&\multicolumn{4}{|c||}{}&\multicolumn{4}{|c|}{}\\
\multicolumn{1}{c}{}&\multicolumn{4}{|c||}{The Minimum $\chi^{2}$ Method}&\multicolumn{4}{|c|}{ $\chi^{2}$ test with scaling factors}\\
\hline
\multicolumn{1}{|c|}{}&\multicolumn{1}{|c|}{}&\multicolumn{1}{|c|}{}&\multicolumn{1}{|c|}{}&\multicolumn{1}{|c||}{ }&\multicolumn{1}{|c|}{}&\multicolumn{1}{|c|}{}&\multicolumn{1}{|c|}{}&\multicolumn{1}{|c|}{}\\
\multicolumn{1}{|c|}{Region}&\multicolumn{1}{|c|}{$|b|<10$}&\multicolumn{1}{|c|}{10$<|b|<$30}&\multicolumn{1}{|c|}{$30<|b|$}&\multicolumn{1}{|c||}{All sky}&\multicolumn{1}{|c|}{$|b|<10$}&\multicolumn{1}{|c|}{10$<|b|<$30}&\multicolumn{1}{|c|}{$30<|b|$}&\multicolumn{1}{|c|}{All sky}\\
\hline
 &  &   &  &   &  & & & \\
  old model     & 5546 & 4184 & 2410 & 11544 & 928 & 863 & 523 & 2231 \\
\hline
                &      &      &      &       &     & & & \\
  Model A       & 358  & 208  & 117  &  399  & 129 & 42 & 28 & 156 \\
\hline
                &      &      &      &       &     &  & & \\
 Model B        & 957  & 162  & 104  &  859  &  197  & 81 & 38 & 345 \\
\hline
                &      &      &      &       &     &  & & \\
 \cite{JJ2010}  & 1230 & 337  & 129  & 1317      & 124 & 94 & 48 & 256 \\
\hline
\end{tabular}
\end{center}
\caption{ $\chi^{2}$ statistics of colour histograms for the old model, two 
default models (A and B), and the \cite{JJ2010} best fit for IMF and SFR with respect to the Tycho-2 histogram in 
different latitude ranges. The two specific tests are explained in the text.}
\label{tableChi}
\end{table*}

\subsection{Deep star counts at the Galactic pole}\label{sec:deepCounts}

\begin{figure*}
\begin{center}
\includegraphics[width=13cm]{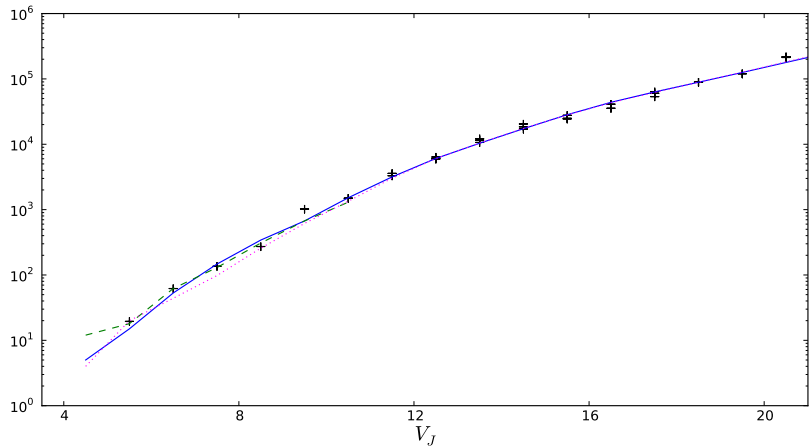}\\ 
\end{center}
\caption{Star count predictions in the V band at the Galactic north pole (  $ |b| > $ 80). We show the 
 original Tycho-2 stars, which are used in our analysis ($V_T \leq $ 11 mag) (green dahsed line) and the simulations
obtained with model A (dotted magenta line) and model B (blue solid line). Black crosses indicate various observed counts 
taken from CDS (5 $< V <$ 8 mag), 
\cite{Gilmore} (12 $< V <$ 18 mag), \cite{Bok} (8 $< V < $ 16 mag), \cite{Yoshii} 
(10 $< V < $17 mag), and Cr\'ez\'e (private communication) (17 $< V <$ 21 mag). }
\label{poles}
\end{figure*}

As the new models improve the fit at magnitudes up to 11 it does not guarantee that they perform well at larger magnitudes.
 Figure \ref{poles} shows a preliminary test made to check the 
performance of these models A and B at a magnitude up to 21 towards the Galactic pole. Simulated star counts are compared with various data sets from CDS (5 $< V <$ 8 mag), 
\cite{Gilmore} (12 $< V <$ 18 mag), \cite{Bok} (8 $< V < $ 16 mag), \cite{Yoshii} 
(10 $< V < $17 mag), and Creze (private communication) (17 $< V <$ 21 mag).
 Both models perform well at deep magnitudes.
 Most of the stars with the apparent magnitude between 10 and 17 have the absolute magnitude within the 
 range 4-10. As expected from their LF presented in Fig. \ref{rhoObs3} and \ref{newModel2}, these models do not differ much in this range.

\section{Conclusions and future work}\label{summary}

We have designed, implemented, and tested a new version of the Besan\c{c}on Galaxy Model.
The new code implements an important change in the star production philosophy of the thin disc 
population, where the IMF, SFR, and stellar evolutionary tracks are treated as free parameters. 
The new code can be tested by comparing simulations by assuming different IMF and SFR, which could also be different 
in different parts of the Galaxy. The code also produces binary systems that take into account the observational spatial
resolution. 

We have updated and analysed several of the model's inputs (see Table \ref{tableParam}) using two observational 
sets, the Tycho-2 
data, and the local luminosity function.  The extinction model is one of the most  crucial 
inputs at low latitudes. Significant differences in the 
Galactic plane are observed 
when using \cite{drimmel2001} or \cite{marshall3D2006} extinction models. Our results indicate that 
\cite{drimmel2001} overestimates the extinction at large distances and \cite{marshall3D2006} underestimates the 
extinction at very short distances. In absence of a better alternative, we 
have selected the \cite{marshall3D2006} whenever it is available and the \cite{drimmel2001} for the rest of the sky.

One of the most crucial parts of our investigation was to look for the best IMF and SFR, which are able to reproduce the 
whole sky Tycho-2 data. When a constant SFR  is assumed, no matter which IMF is used, we are not able to reproduce 
the Tycho-2 star counts. The model always shows a significant 
excess of blue stars. This result strongly favours a decreasing SFR. 
 The new decreasing SFR adopted here does not alter the overall good agreement obtained with the previous model
(constant SFR) at fainter magnitudes. This is due to the fact that fainter stars (V$>$14) of the thin disc are on the
 lower main sequence where the stellar density is sensitive to the integral of the SFR over the disc age and not to the 
detailed evolution in time.

When looking for the best fit IMF, we notice the following: 

\begin{itemize}
\item A slope $\alpha$=2.3 at high masses, which has been used by \cite{Vallenari2006} and \cite{Kroupa2008} produces a 
strong excess of blue stars. As already mentioned by \cite{misha1997-1}, a Salpeter IMF is certainly inappropriate
for our Galaxy. A steeper slope in this mass range is recommended.
\item  The \cite{JJ2010} IMF used with their best fit decreasing SFR, and our default model reproduces the Tycho-2 
$(B-V)_T$ distribution at intermediate and high latitudes well, but it performs a bit worse within the plane.
\end{itemize}

We conclude that Kroupa-Haywood v6 and Haywood-Robin IMF combined with the decreasing SFR 
of \cite{AumerBinney2009} provide the best fit to Tycho data. 

All the thin disc ingredients cannot be constrained using nearby photometric data alone. A better solution would 
require a higher amount of observational constraints (in particular deeper star counts) and a robust statistical model
fitting algorithm
(as for instance Markov Chain Monte Carlo method). Nevertheless, we have learned that Tycho data is able to impose strong observational
constraints to the SFR and the IMF in the bright star/intermediate mass domain in the solar vicinity. 

A complementary investigation of the thick disc is underway using SDSS and 2MASS surveys  
to characterise the thick disc properties and constrain the scenario of its formation. We believe that using this new 
scheme and a better knowledge of the thin disc itself will help to perform this analysis, getting rid of the uncertainties 
on the structural parameters due to the degeneracies of parameters and to the thin disc contamination. Scenarios, such as 
the decomposition of populations of mono-abundances from \cite{Bovy}, the question of the role of migration, in situ 
formation, or formation by mergers will then be possibly tested.

The next step will be to test the stellar kinematics and metallicity link to constrain the chemo-dynamical evolution
of the Galaxy using RAVE data. The availability of even deeper spectroscopic probes, like the Gaia-ESO survey or APOGEE, 
will also allow us to 
constrain the SFR on a larger scale and to specifically study the link between the local thin and thick discs and 
more problematic populations of the bar and the bulge in the inner Galaxy. We expect that the new tool presented here
will be very 
efficient for the analysis and understanding of future large scale surveys, such as Gaia.
Work is in progress to 
make the new default models available online.

\begin{acknowledgements}
We thank O. Bienaym\'e and M. Cr\'ez\'e for useful discussions and suggestions. We also wish to thank to F. Arenou for providing us the detailed strategy for
binarity implementation. The anonymous referee is thanked for proposing improvements to clarify the paper. M. A. Czekaj was supported as an Early Stage Researcher of the Marie Curie Research Training 
Network "European Leadership in Space Astrometry" (ELSA) MRTN-CT-2006-033481 of the VIth 
Framework Programm - European Community and the MICINN (Spanish Ministry of Science and 
Innovation) - FEDER through grant AYA2009-14648-C02-01. We acknowledge the support of the 
French 'Agence Nationale de la Recherche' under contract ANR-2010-BLAN-0508-01OTP. This thesis 
has been carried out at the Department d'Astronomia i Meteorologia (Universitat de Barcelona)
 and the Observatoire de Besan\c{c}on (France). This study was partially supported by the Gaia 
Research for European Astronomy Training (GREAT) 08-RNP-118 European Science Foundation (European
 RNP FP7) and the MICINN - AYA2009-08488-E/AYA.
\end{acknowledgements}

\bibliographystyle{aa}
\bibliography{ref}

\appendix

\section{Statistical treatments}

\subsection{Small volume elements treatment}
\label{volume}

\begin{figure} 
\begin{center}
\includegraphics[width=9cm]{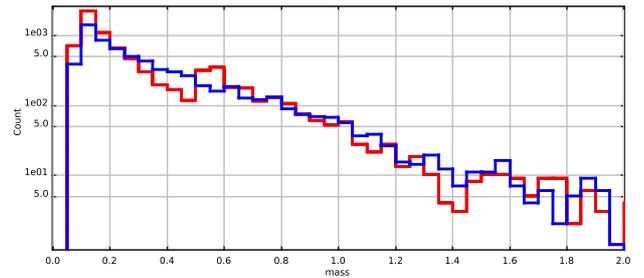}\\
\end{center}
\caption{ Simulated mass distributions. Three slope IMF was assumed in these
simulations: $\alpha_1$=1.3 for 0.09 $< m <$ 0.5 $M_{\odot}$,  
$\alpha_2$=2.3 for 0.5 $< m <$ 1.53 $M_{\odot}$ and $\alpha_3$=3.0 for 1.53 $< m <$ 120 $M_{\odot}$. In red the 
sample biased due to the problem of small volume elements and in blue the simulations obtained after 
the implementation of small volume elements treatment (the correcting factor was set to 50). }
\label{IMFtesting}
\end{figure}

A simple test was performed to demonstrate the problem due to the use of small volume elements at short heliocentric
distances. We have performed the simulations of a
standard region (40 square deg) until 100 pc. When no additional treatment is applied in the volume elements, the 
resulting mass distribution presents significant discontinuities (see Fig. \ref{IMFtesting}).
They are caused by too low mass enclosed inside the volume being processed. We checked that this effect is 
less significant when no cut in distance is applied, because the further volume elements with more mass
smooth out the dip. To correct this effect, the code is able to enlarge the volume element by a given factor and draw the masses from
that enlarged pool. Through fitting, we estimated that a factor of 50 is enough to avoid the bias. This strategy 
assures that 
the mass calculated for that enlarged volume 
element is big enough for different masses to be drawn with no bias. Later on, we (randomly) keep only a fraction of the drawn objects, 
which corresponds to the original small volume element. As shown in Fig. \ref{IMFtesting}, this treatment avoids the 
underestimation of the number of stars from the high-mass IMF tail and leads to no discontinuities in the mass 
distribution of the simulated sample.

\subsection{What is the IMF of secondary stars?}
\label{IMFsec}

\begin{figure}
\begin{center}
\includegraphics[width=8cm]{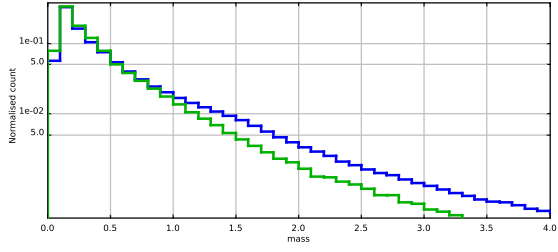}\\
\end{center}
\caption{ The difference in the IMF of single (blue) and secondary (green) stars for the full sky sample (r$<$100 pc).}
\label{imf1}
\end{figure}

\begin{figure}
\begin{center}
\includegraphics[width=8cm]{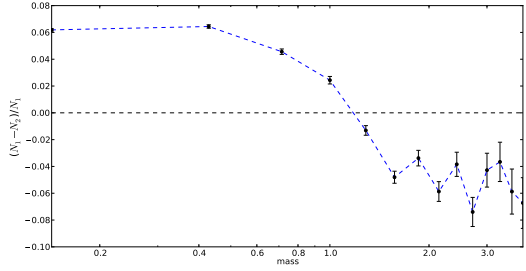}\\
\end{center}
\caption{Relative difference of the number of objects when using the binarity treatment $N_1$ containing single, primary and secondary stars, and the sample without binaries $N_2$ that is the IMF of single stars. }
\label{imf2}
\end{figure}

As discussed is Section \ref{subsec:binarityImplementation}, we have preferred to produce the masses of the 
secondary stars from the empirical relations \citep{Arenou2011} and not from the IMF that is assumed for single and primary stars.
To verify the differences between both IMFs, we 
 performed the simulations of the sphere around the Sun (100 and 180 pc) with and without binaries. In both cases, we 
 saved the masses of all ever produced stars (alive and remnants). 
In Fig. \ref{imf1}, we compare the mass distribution of primaries and secondary stars. 
In Fig. \ref{imf2}, we show the relative difference of the number of objects when using the binarity treatment $N_1$
that contains single, primary, and secondary stars and the sample 
without binaries $N_2$ (that is the IMF of single stars). We only plot  the mass 
range m=[0:4] $M_\odot$ because of the low statistics at higher masses. There is a difference at the level of 6 \% due
to secondary stars. Thus, in conclusion, the IMF is not preserved. The IMF of secondary stars is not 
 known enough, and it has no strong reason to be the same as the single star IMF. 
 We decide to keep the constraint from  the statistics of observed binaries, rather than the  unverified theroretical
 hypothesis that the IMF of secondaries is the same as the primaries and singles.

\section{Photometry transformations}
\label{photo}

To find the best transformation between the Johnson and Tycho systems, we have analysed four different approaches.
First, we considered the standard transformation published in the Vol. 1 of \cite{ESA1997} (see Sections 1.3 and 2.2)

\begin{displaymath}
 V_{J} = V_{T} - 0.090(B_{T} - V_{T}), 
\end{displaymath}

\begin{equation}
 B_{J} - V_{J} =  0.850(B_{T} - V_{T}).
\label{std}
\end{equation}

The second approach was the linear interpolation of the values, as specified in Table 1.3.4. of the same publication. 
In this case, the slope of the $(B-V)_{J}$ versus $(B_{T} - V_{T})$ relation, or the so called G-factor, is different in each 
of the six presented colour intervals. 

In a third approach, we used equations of giant-like stars (luminosity class III with 
low reddening) for the whole sample (also from Vol. 1 of \cite{ESA1997}). 

The fourth method of transformation comes from 
Mark Kidger. On his website\footnote{http://www.britastro.org/asteroids/Tycho\%20Photometry.htm}, he derives the following
relations:

\begin{displaymath}
(B_T-V_T) = 1.28899 (B_{J}-V_{J}) - 0.1031 \\
\end{displaymath}

\begin{equation}
V_J = V_T - 0.016 - 0.0741 \times (B_T-V_T).
\end{equation}

\begin{figure} [ht!]
\begin{center}
\includegraphics[width=9cm]{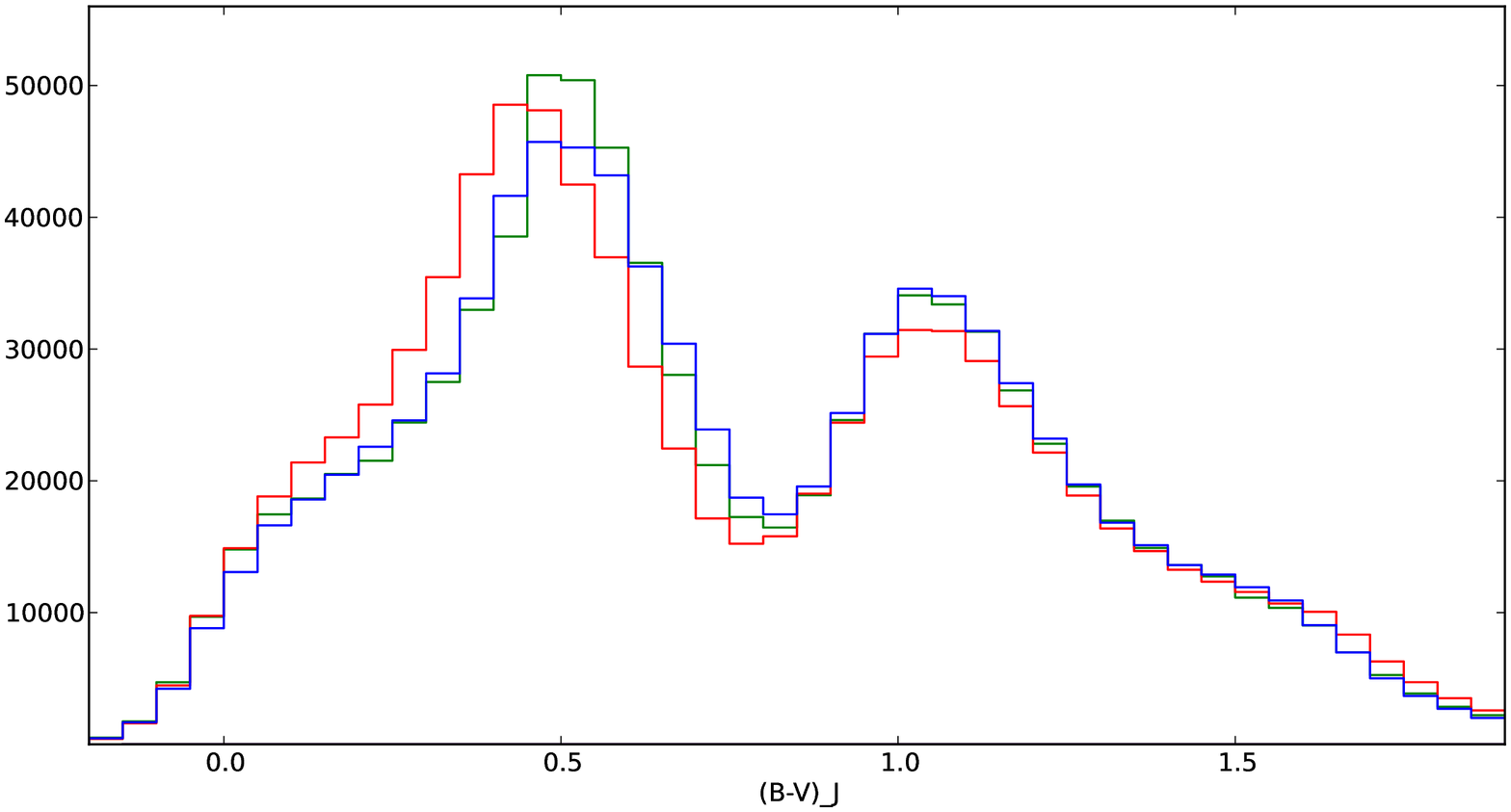}\\
\includegraphics[width=9cm]{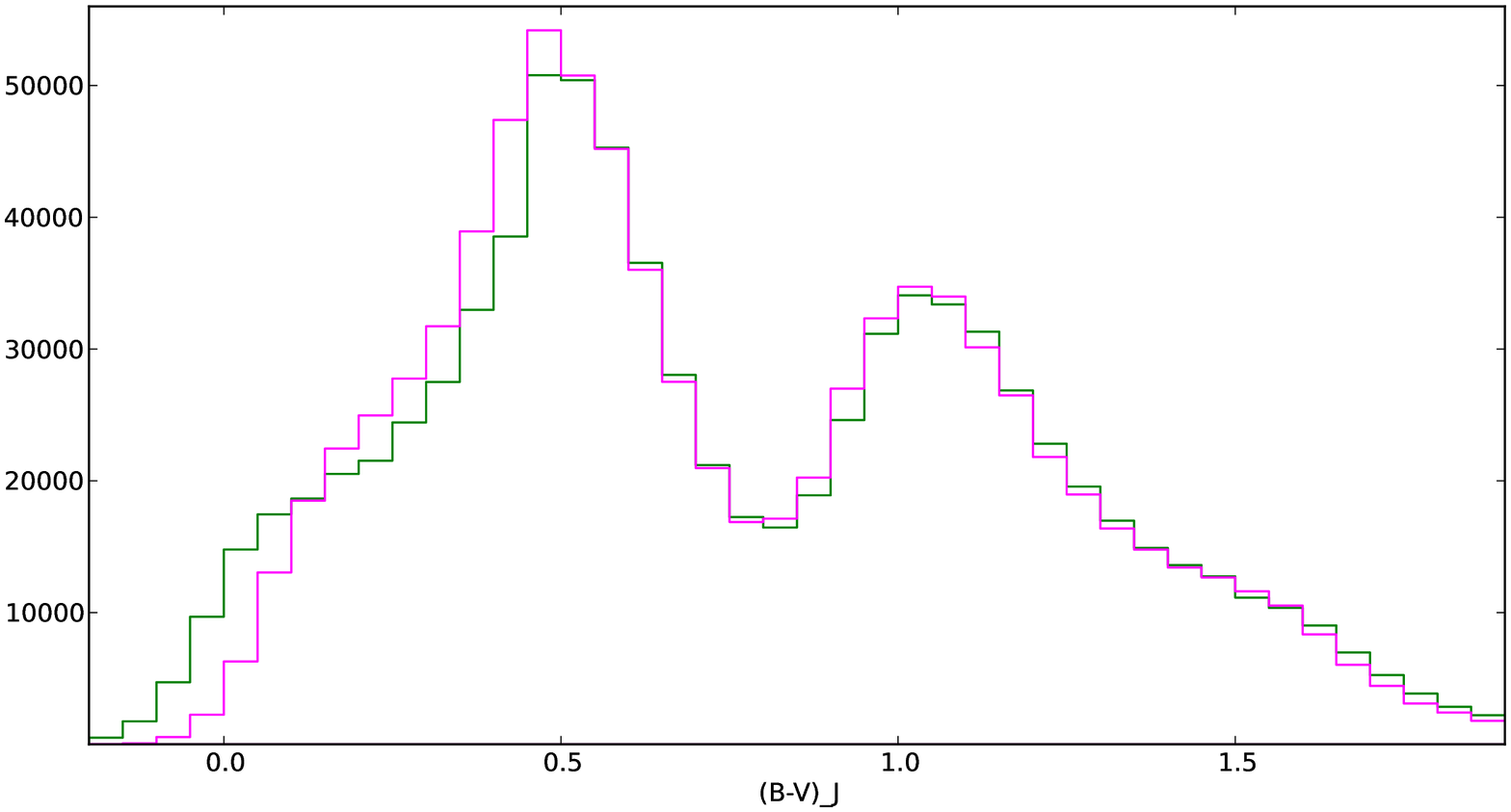}\\
\end{center}
\caption{$(B-V)_J$ distribution of Tycho-2 catalogue (after cutting at $V_T$ = 11 mag) transformed using four different 
photometry transformations. Top: Standard (red),linear interpolation (green), and giants (blue) methods. Bottom: Linear 
interpolation (green) and Mark Kidger's (magenta) methods.}
\label{p1}
\end{figure}

In Fig. \ref{p1}, we present the $(B-V)_J$ distributions of the Tycho sample (cut at $V_T<=$ 11 
mag) transformed into a Johnson system by the four discussed methods. Both the standard transformation, and the 
relation derived by Mark Kidger 
propose a unique slope for all the colour ranges. These standard equations are a rough approximation because they 
impose the same transformation for all types of stars. \cite{perryman1997hipparcos} suggest a more sophisticated method of transformation such as the linear interpolation approach. 
As expected, the giants and linear interpolation methods show very similar results for the giants peak, while significant
 differences appear in the blue peak. 
In this paper, we use the linear interpolation
transformation method, so we have inverted it and transformed the photometry of our simulations from Johnson to Tycho-2 system.

\end{document}